\documentclass{ws-ijmpd}
\usepackage[super,compress]{cite}
\usepackage{amsmath}
\usepackage{mathrsfs}
\usepackage{booktabs}
\usepackage{float}
\usepackage{placeins}
\usepackage{pdflscape}
\usepackage{pdfpages}

\begin{document}

\markboth{M. Lawrence Pattersons, Freddy P. Zen, Hadyan L. Prihadi, Muhammad F. A. R. Sakti, Getbogi Hikmawan}
{Rotating neutron stars with chaotic magnetic fields in general relativity and Rastall gravity}

%
\catchline{}{}{}{}{}
%

\title{ROTATING NEUTRON STARS WITH CHAOTIC MAGNETIC FIELDS IN GENERAL RELATIVITY AND RASTALL GRAVITY
}

\author{M. LAWRENCE PATTERSONS\footnote{m.pattersons@proton.me}}

\address{Theoretical High Energy Physics Group, Department of Physics, Institut Teknologi Bandung\\
Jl. Ganesha 10, Bandung 40132,
Indonesia}

\author{FREDDY P. ZEN\footnote{fpzen@fi.itb.ac.id}}

\address{Theoretical High Energy Physics Group, Department of Physics, Institut Teknologi Bandung\\
Jl. Ganesha 10, Bandung 40132,
Indonesia}
\address{Indonesia Center for Theoretical and Mathematical Physics (ICTMP), Institut Teknologi Bandung\\
Jl. Ganesha 10, Bandung 40132,
Indonesia}

\author{HADYAN L. PRIHADI\footnote{hady001@brin.go.id}}

\address{Research Center for Quantum Physics, National Research and Innovation Agency (BRIN)\\
South Tangerang 15314,
Indonesia}

\author{MUHAMMAD F. A. R. SAKTI\footnote{fitrahalfian@gmail.com}}

\address{High Energy Physics Theory Group, Department of Physics, Faculty of Science, Chulalongkorn University\\
Bangkok 10330,
Thailand}

\author{GETBOGI HIKMAWAN\footnote{getbogihikmawan@itb.ac.id}}

\address{Theoretical High Energy Physics Group, Department of Physics, Institut Teknologi Bandung\\
Jl. Ganesha 10, Bandung 40132,
Indonesia}
\address{Indonesia Center for Theoretical and Mathematical Physics (ICTMP), Institut Teknologi Bandung\\
Jl. Ganesha 10, Bandung 40132,
Indonesia}

\maketitle

\begin{history}
\received{Day Month Year}
\revised{Day Month Year}
\end{history}

\begin{abstract}
Observations indicate that the magnetic fields on neutron stars (NSs) lie in the range of $10^{8}$–$10^{15}$~G. We investigate rotating NSs with chaotic magnetic fields in both general relativity (GR) and Rastall gravity (RG). The equation of state (EOS) of NS matter is formulated within the framework of quantum hadrodynamics (QHD). The Hartle–Thorne formalism, extended to RG, is employed as an approximation for describing rotating NSs, while the magnetic field is modeled through an ansatz in which it is coupled to the energy density. We find that at high masses, neither rotation nor the Rastall parameter significantly affects the total mass, whereas the magnetic field strength can increase the maximum allowed mass. At lower masses, both the magnetic field and an increasing Rastall parameter reduce the stellar radius in the static configuration. Although higher angular velocities enhance stellar deformation, both magnetic field and larger Rastall parameter tend to suppress it. Regarding the moment of inertia, the Rastall parameter has little impact, whereas the magnetic field strength can increase it within the mass range $1.50$–$1.99~M_\odot$. All parameters considered in this study are consistent with observational constraints on the moment of inertia obtained from radio observations of massive pulsars.
\end{abstract}


\ccode{PACS numbers:}


\section{Introduction}
\label{sec1}

\label{sec1}
It has been widely understood that the evolution of the massive stars whose mass is 8$-$25 $M_\odot$ ends when neutron stars (NSs) form. The corresponding formation is due to the stars’ gravitational collapse during Type-II, Ib, or Ic supernova explosion phenomena~\cite{Pattersons2021,Vidaña2018}. NSs are compact remnants of evolved stars, with degenerate fermions compressed by strong gravitational fields \cite{Reisenegger2001}. They are compact enough to provide necessary conditions for exotic physics to occur \cite{Abac2023}. The stabilizing effect of gravity allows long-time-scale weak interactions (such as electron captures) to reach equilibrium, forming matter that is neutron-rich \cite{Graber2017}. These stars cover an extensive density range, roughly several times of nuclear density ($n_0 \sim 10^{14}$ gm/cc), throughout their outer crust to their inner core \cite{Ghosh2024}. NSs are normally rotating as all celestial objects tend to naturally always rotate due to the law of angular momentum conservation \cite{Sivaram2012}. When rotating NSs are highly magnetized, they are recognized as magnetars \cite{Cohen1972,Horvath2004,Lyne2013,Basu2018,Fortin2021}.

NSs have a wide range of magnetic fields. Observations indicate that magnetic fields on NSs are at least in the range of $10^{8-15}$ G \cite{Reisenegger2001}. It is important to note that the direct observation of the magnetic field in NSs' cores remains unachieved. However, theoretical prediction suggests that the strengths of the magnetic fields in NSs' cores are in the order of $10^{18}$ G to $10^{20}$ G \cite{Rizaldy2018b,Franzon2016}. New classes of pulsars such as Anomalous X-ray pulsars (AXPs) and Soft-Gamma Repeaters (SGRs) have been identified producing vast magnetic fields. SGR is associated with remnants of a supernova, which is a young NS \cite{Rizaldy2018a}.

Highly magnetized NSs, i.e. magnetars, are typically slowly rotating NSs \cite{Desvignes2024}. The recognized fastest-spinning magnetar is Swift J1818.0--1607 \cite{Ding2024}, whose period is only 1.36 s \cite{Fisher2024}. Magnetars' emissions are thought to be powered by the decay of their large ($\geq 10^{12}$ G) magnetic fields, much greater than pulsars \cite{Desvignes2024}, even surpassing the magnitude of the QED critical field \cite{Gau2024}. Moreover, observations on some AXPs also indicate that magnetic field of AXPs surface is around $10^{14}-10^{15}$ G \cite{Rizaldy2018a}. 

Konno et al. [\citen{Konno2000}] calculated the deformation of NSs resulting from rotation and the presence of magnetic fields. In their analysis, a polytropic equation of state (EOS) was employed. Their findings indicate that for magnetars, the deformation induced by magnetic fields is dominant, whereas for typical pulsars, the magnetic effect is negligible. Mallick \& Schramm [\citen{Mallick2014}] investigated mass corrections and the deformation of NSs under the influence of magnetic fields, treating the magnetic field as an anisotropy in the energy-momentum tensor (EMT). Interestingly, despite focusing on static NSs, their approach was based on the Hartle-Thorne (HT) formalism [\citen{Hartle1967,Hartle1968}], which is typically applied to rotating relativistic stars. In their work, magnetic perturbations in static configurations were treated analogously to rotational perturbations, following the HT approximation. They considered both very stiff and very soft EOS models in their study. Lopes \& Menezes [\citen{Lopes2015}] explored the effects of chaotic magnetic fields on NSs. A key advantage of incorporating a chaotic magnetic field lies in the elimination of anisotropy, simplifying the mathematical formulation. They also introduced an ansatz where the magnetic field is coupled to the energy density of the NS. For their EOS models, they utilized two variations: one including hyperons and one without hyperons. However, the rotation of NSs was not considered in their analysis.

An important property of NSs is their moment of inertia, which offers insights into their internal structure and EOS. Given the link between the EOS and moment of inertia, approximate methods for estimating it in static models, particularly for stiff EOS scenarios, are feasible \cite{daSilva2021}. Furthermore, measuring the moment of inertia is crucial due to its universal correlation with compactness \cite{Rahmansyah2020}. Rahmansyah et al. [\citen{Rahmansyah2020}] calculated the moment of inertia of anisotropic NS which satisfies the constraints proposed by Landry et al. [\citen{Landry2020}]. However, the magnetic field is not taken into account in their calculation of the moment of inertia.

The works that has been mention above are constructed upon general relativity (GR). Moreover, the works presented in Refs.~[\citen{Frieben2012,Pili2014,Pili2017}], which deal with rotating NSs with magnetic fields, are also based on GR. However, if GR is correct, there are some problems left to be explained, e.g. dark matter and dark energy \cite{Clifton2012,Apryandi2025}. Dark energy cannot be consistently described solely by GR. The reason is that the dark energy EOS $\omega_{DE}$ is constrained to have values $\omega_{DE}=-1.018\pm 0.031$. This fact means that it is a probability that dark energy is actually phantom dark energy. In the context of GR, phantom dark energy is described only by phantom scalars, which do not provide a physically acceptable description of nature \cite{Odintsov2021}. In addition, the Hubble tension—referring to the discrepancy in the measurements of the rate of the expansion of the universe, when comparing local measurements with cosmic microwave background radiation observations —remains another unresolved issue within the framework of GR \cite{Fortunato2025}. There are also some problems related to compact objects. For example, the problem of the maximum allowed mass of horizonless compact objects exceeded GR predictions \cite{Danarianto2025}. As the consequence, one might consider modified theories of gravity as the alternative ways. One of them is Rastall gravity (RG) [\citen{Rastall1972}].

Unlike GR, RG allows for a non-vanishing covariant divergence of the energy-momentum tensor (EMT), with the divergence being proportional to the gradient of the Ricci scalar \cite{Oliveira2015}. Thus, RG is claimed as a generalization of GR \cite{Rastall1972}. However, the physical interpretation of the additional source term in RG remains non-trivial. Phenomenologically, it can be regarded as an effective manifestation of quantum effects within a classical context \cite{Oliveira2015}. Although controversial and claimed to be equivalent to GR \cite{Visser2018}, many studies [\citen{daSilva2021,Oliveira2015,Darabi2018,Sak2,Men,Tan,Pre,Nas2,Nas,Tan2,Sal,Xi,Pattersons2024}] have demonstrated that RG is indeed different from GR. Furthermore, RG has been extensively applied to the study of various compact astrophysical objects, including black holes (BHs) \cite{Sak2,Nas,Sak,Pri,Spa,Hey,Zou,Kumar,Heydar}, wormholes \cite{Bha,Malik,Hal,Mor,Hey2,Bro}, strange quark stars \cite{Sal,Aya}, and gravastars \cite{Maj}. In the cosmological context, RG has also been employed to study accelerated expanding universe \cite{Cap}. A specific refutation of the claimed equivalence between RG and GR can be found in Ref.~[\citen{Darabi2018}].

{Especially in theoretical astrophysics, several studies have reported notable findings by incorporating RG. Pattersons et al. [\citen{Pattersons2024}] showed that although RG only slightly increases the mass of anisotropic NSs, the resulting stellar mass can still reach the upper bound of the mass range for the secondary compact object of GW190814 through the contribution of the Rastall parameter. It is important to note that the secondary component of GW190814 \cite{Abbott2019} has been hypothesized to be the most massive NS ever observed \cite{Fattoyev2020}. Consistently, da Silva et al. [\citen{daSilva2021}] found that NSs in RG can attain this upper limit, while Banerjee et al. [\citen{Banerjee2024}] demonstrated that quark stars within RG may also reach it.}

As previously mentioned, the central idea of RG lies in the proportionality between the covariant divergence of the EMT and the gradient of the Ricci scalar. Consequently, the essence of RG is inherently linked to high-curvature environments, making NSs promising natural laboratories for testing its predictions \cite{Oliveira2015}. Several works have involved NSs in RG. Oliveira et al. [\citen{Oliveira2015}] formulated the structure equations for spherically symmetric isotropic NSs and obtained their numerical solutions. Xi et al. [\citen{Xi}] investigated the same case using a different EOS. Da Silva et al. [\citen{daSilva2021}] extended the formulation to include rotating isotropic NSs. Meng and Liu [\citen{Men}] derived the tidal Love numbers of NSs. Majeed et al. [\citen{Maj2}] developed models of static anisotropic compact stars. Pattersons et al. [\citen{Pattersons2024}] obtained the rotational mass of anisotropic NSs. It is worth noting that Ref.~[\citen{daSilva2021}] and Ref.~[\citen{Pattersons2024}] worked on rotating NSs in RG. However, the deformation of the NSs due to rotation is not considered.

Building on previous studies, we investigate the physical properties of rotating NSs with a chaotic magnetic field in GR and RG, focusing on the mass–radius relation, eccentricity, and moment of inertia. The magnetic field is modeled using the Lopes–Menezes ansatz, in which the magnetic field strength is coupled to the energy density of the stellar matter. The rotational aspects are modeled using the HT formalism adapted to the RG framework. In this work, we present the formulation of stellar deformation within the rotational configuration in RG, which was not considered in Refs.~[\citen{daSilva2021,Pattersons2024}]. As a result, the mass–radius relation becomes more accurate by taking into account the deformation of the star. Our results demonstrate that the free parameter of RG significantly affects the deformation behavior of NSs. This supports the relevance of studying stellar deformation within the RG framework. 

For the analysis of the moment of inertia, we adopt the very slow rotation approximation with a fixed angular velocity of $\Omega = 50$ s$^{-1}$, corresponding to a nearly spherical configuration assumed for simplicity. In this regime, stellar deformation is negligible and the eccentricity approaches zero. This justifies the use of the standard moment of inertia formula for a spherical object. A similar approach has also been employed in Refs.~[\citen{Rahmansyah2020,Rizaldy2024}].

In general, NS matter is described by its EOS. In this work, we employ the quantum hadrodynamics (QHD) EOS as used in Ref.~[\citen{Lopes2020}]. The QHD framework treats baryons as the fundamental degrees of freedom and models their interactions via meson exchange. The EOS provides the pressure and energy density, which serve as inputs for both the GR and RG calculations. The numerical calculations follow the algorithm used in Refs.~[\citen{Pattersons2024,Ordaz2019}].

The remainder of this paper is organized as follows. In Section 2, we outline the formulations employed in this study. Section 3 presents the numerical results and a discussion of their implications. Finally, we provide a summary of our findings in Section 4.

\section{Mathematical Formulations}
To make the discussion self-contained, we briefly present the nuclear model of the EOS employed in this work in Sect. 2.1. In Sect. 2.2, we review the chaotic magnetic field structure of NSs, followed by a brief overview of RG in Sect. 2.3. The HT formalism within the framework of RG is presented in Sect. 2.4. Throughout the mathematical formulations, we adopt geometrized units by setting $G=c=1$.
\subsection{{Nuclear Model of the EOS}}
All mathematical formulations regarding the EOS used in this work are based on Ref.~[\citen{Lopes2020}]. The QHD Lagrangian is given by
\begin{eqnarray}
\mathscr{L}_{\textsf{QHD}} &=& \sum_b \bar{\psi}_b \left[ \gamma^\mu \left( i\partial_\mu - e_b A_\mu 
- g_{b,\omega} \omega_\mu - g_{b,\rho} \frac{1}{2} \boldsymbol{\tau} \cdot \boldsymbol{\rho}_\mu \right) 
- (m_b - g_{b,\sigma} \sigma) \right] \psi_b \nonumber \\
&& + \frac{1}{2} m_\omega^2 \omega_\mu \omega^\mu 
+ \frac{1}{2} m_\rho^2 \boldsymbol{\rho}_\mu \cdot \boldsymbol{\rho}^\mu 
+ \frac{1}{2} \left( \partial_\mu \sigma \partial^\mu \sigma - m_\sigma^2 \sigma^2 \right) 
- U(\sigma) \nonumber \\
&& - \frac{1}{4} F^{\mu\nu} F_{\mu\nu} 
- \frac{1}{4} \Omega^{\mu\nu} \Omega_{\mu\nu} 
- \frac{1}{4} \mathbf{P}^{\mu\nu} \cdot \mathbf{P}_{\mu\nu}.
\label{lagrangian}
\end{eqnarray}
Here $\psi_b$ are the baryonic Dirac fields; $\sigma$, $\omega_\mu$, $\rho_\mu$ denote the mesonic fields; The $g$'s represent the Yukawa coupling constants, which simulate the strong interaction; $m_b$ is mass of baryon $b$; $m_\sigma$, $m_\omega$, and $m_\rho$ denote the masses of $\sigma$, $\omega$, and $\rho$ mesons, respectively; $e_b$ represents the electric charge of baryon $b$; $A_\mu$ denotes the electromagnetic four-potential; $\boldsymbol{\tau}$ are the Pauli matrices. The antisymmetric mesonic fields' strength tensors are given by
\begin{eqnarray}
    &F^{\mu\nu}=\partial^\mu A^\nu-\partial^\nu A^\mu,&\\
    &\Omega^{\mu\nu}=\partial^\mu \omega^\nu-\partial^\nu\omega^\mu,&\\
    &\boldsymbol{P}_{\mu\nu}=(\partial_\mu\overrightarrow{\rho}_\nu-\partial_\nu\overrightarrow{\rho}_\mu)-g_\rho(\overrightarrow{\rho}_\mu\times\overrightarrow{\rho}_\nu).&
\end{eqnarray}
The term $U(\sigma)$ represents the self-interaction term that is used to reproduce some saturation properties of the nuclear matter, and is given by
\begin{eqnarray}
    U(\sigma)=\frac{1}{3!}\kappa\sigma^3+\frac{1}{4!}\lambda\sigma^4.
\end{eqnarray}

The Lagrangian of leptons writes
\begin{eqnarray}
    \mathscr{L}_{\text{lep}}=\sum_l \bar{\psi}_l [i\gamma^\mu(\partial_\mu -e_l A_\mu)-m_l]\psi_l.
\end{eqnarray}
Here the sum runs over $e$ and $\mu$.

\begin{table}[ht]
    \centering
    \caption{GM1 model parameters and physical quantities, the values are taken from Ref.~[\citen{Lopes2020}], in which the phenomenological constraints are taken from Refs.~[\citen{Oertel2017,Glendenning1991,Dutra2014}].}
    \label{tab:example}
    \begin{tabular}{lccccc}
        \toprule
                  & Parameters &   & Phenomenology & GM1 & Masses (MeV)  \\ 
        \midrule
        $(g_{N\sigma}/m_\sigma)^2$           & 11.785 fm$^2$   & $n_0$ (fm$^{-3}$) & $0.148-0.170$ & 0.153 & $M_\Lambda=1116$    \\
        $(g_{N\omega}/m_\omega)^2$           & 7.148 fm$^2$   & $M^*/M$ & $0.7-0.8$ & 0.7 & $M_\Sigma=1193$    \\
        $(g_{N\rho}/m_\rho)^2$           & 3.880 fm$^2$   & $K$ ($-$MeV) & $200-315$ & 300 & $M_\Xi=1318$    \\
        $\kappa/M_N$           & $0.005894$   & $S_0$ (MeV) & $30-34$ & 30.5 & $m_e=0.511$    \\
        $\lambda$ & $-0.006426$ & $B/A$ (MeV) & $15.7-16.5$ & 16.3 & $m_\mu = 105.6$ \\
        $M_N$ & $939$ MeV & $L$ (MeV) & $36-113$ & 88 & $-$ \\
        \bottomrule
    \end{tabular} \label{parameterset}
\end{table}

The EOS used in this work is based on the GM1 parameterization, which is also used in Ref.~[\citen{Lopes2020}]. There are six considered properties at the saturation density, i.e. the saturation density itself $n_0$, the effective nucleon mass $M/M^*$, the compressibility $K$, the symmetry energy $S_0$, the binding energy per baryon $B/A$, and the slope of the symmetry
energy $L$. The prediction of the physical quantities and their inferred values from phenomenology is summarized in Table \ref{parameterset}.

According to Ref.~[\citen{Lopes2020}], hyperons are included in this EOS model. The potential depth is fixed using the well-known $\Lambda$ potential depth, $U_\Lambda = -28~\text{MeV}$. The hybrid SU(6) symmetry group is employed to fix all vector mesons, while a nearly SU(6) symmetry is also used for the scalar mesons (see also Refs.~[\citen{Pais1961,Lopes2014}]). All hyperon-meson coupling constants are given below in Eq.~\ref{couplingcontants}

\begin{eqnarray}
    &\frac{g_{\Lambda\omega}}{g_{N\omega}}=\frac{g_{\Sigma\omega}}{g_{N\omega}}=0.667,\:\:\frac{g_{Xi\omega}}{g_{N\omega}}=0.333,&\nonumber\\
    &\frac{g_\Sigma\rho}{g_{N\rho}}=2.0,\:\:\frac{g_{\Xi\rho}}{g_{N\rho}}=1.0,\:\:\frac{g_{\Lambda\rho}}{g_{N\rho}}=0.0,&\nonumber\\
    &\frac{g_{\Lambda\sigma}}{g_{N\sigma}}=0.610,\:\:\frac{g_{\Sigma\sigma}}{g_{N\sigma}}=0.396,\:\:\frac{g_{\Xi\sigma}}{g_{N\sigma}}=0.113.&\label{couplingcontants}
\end{eqnarray}.

The Euler-Lagrange equation for baryons in the absence of electric field gives
\begin{eqnarray}
\left[ \gamma_0 (i \partial^0 - g_{b,\omega} \omega_0 - g_{b,\rho} \rho_0) 
- \gamma_j (i \partial^j - e_b A^j) - M_b^* \right] \Psi &=& 0,
\end{eqnarray}
Here $M_b^* \doteq m_b - g_{b,\sigma} \sigma_0$ is the effective baryon mass.

For uncharged baryons, $e_b A^j$ always vanishes. By using the quantization rules ($E=i\partial^0$, $k=i\partial^j$), the energy eigenvalue of the baryons reads
\begin{eqnarray}
E_b &=& \sqrt{k^2 + M_b^{*2}} + g_{b,\omega} \omega_0 + g_{B,\rho} \frac{\tau_3}{2} \rho_0. \label{eq:Eb1}
\end{eqnarray}

For the case of magnetars, the starting point is a static external magnetic field in the $z$ direction (these conditions will be relaxed later). The potentials are given by $A_2 = A_3 = 0$; $A_1 = -B_0 y$. The eigenvalue writes

\begin{eqnarray}
E_b &=& \sqrt{k_z^2 + M_b^{*2} + 2 \nu |e| B_0} + g_{b,\omega} \omega_0 + g_{B,\rho} \frac{\tau_3}{2} \rho_0, \label{eq:Eb2}
\end{eqnarray}
Here $\nu$ is the Landau level (LL).

For the leptons, we have:
\begin{eqnarray}
E_l &=& \sqrt{k_z^2 + m_l^{2} + 2 \nu |e| B_0}. \label{eq:Elepton}
\end{eqnarray}

The energy density is given by
\begin{eqnarray}
\varepsilon &=& \frac{|e| B_0}{2\pi^2} \sum_{\nu} \eta(\nu) \int_{0}^{k_f} \sqrt{ k_z^2 + M_b^{*2} + 2\nu |e| B_0 } \, dk_z. \label{eq:epsilon}
\end{eqnarray}

Here $k_f$ is the Fermi momentum of the particle and $\eta(\nu)$ is the degeneracy of the LL $\nu$.

The contribution of the mesonic fields to the energy density reads
\begin{eqnarray}
\varepsilon_m &=& \frac{1}{2} \Big( m_\sigma^2 \sigma_0^2 + m_\omega^2 \omega_0^2 + m_\rho^2 \rho_0^2 \Big) + U(\sigma). \label{eq:epsilonm}
\end{eqnarray}
The total energy density is the sum of the energy density of baryons, leptons and mesons.

The EOS is given via thermodynamics
\begin{eqnarray}
    p=\sum_f \mu_f n_f - \varepsilon,
\end{eqnarray}
where $\mu_f$ is the chemical potential, $n$ is the number density, and $f$ runs over all fermions. Please note that more details about the nuclear model of the EOS can be referred to Ref.~[\citen{Lopes2020}].

\subsection{Chaotic Magnetic Field of Neutron Stars}
Generally, chaotically tangled magnetic field lines are thought to be present throughout astrophysical plasmas: planets, stars, accretion discs, galaxies, clusters of galaxies, and the intergalactic medium \cite{Kumar2017}. In nature, chaotic magnetic fields could be generated by asymmetric current configuration \cite{Dasgupta2014}.

It has been widely understood that the magnetic field could generate anisotropy within NSs \cite{Rizaldy2018b,Rizaldy2018a,Mallick2014}. This problem would make the HT formalism for rotational configuration becomes more complex (please see Refs.~[\citen{Pattersons2021,Posada2024,Becerra2024}]). For the magnetic field $B$ in the $z$-direction, it is well-known that the stress tensor is written in the form: $diag(\frac{B^2}{8\pi}; \frac{B^2}{8\pi}; -\frac{B^2}{8\pi})$, being non identical \cite{Lopes2015}. In Ref.~[\citen{Zeldovich}], it is argued that the effect of a magnetic field can be described using the concept of pressure only in the case of a small-scale chaotic field. Under this condition, the pressure due to the magnetic field $p_B$ is shown to be consistent with field theory. Authors in Ref.~[\citen{Lopes2015}] agreed to this argument. Now $p_B$ reads
\begin{equation}
    p_B=\frac{1}{3}<T^j_j>=\frac{1}{3}\left(\frac{B^2}{8\pi}+\frac{B^2}{8\pi}-\frac{B^2}{8\pi}\right)=\frac{B^2}{24\pi},
\end{equation}
where $T^j_j$ denotes the spatial components of the EMT. With this formulation in our hand, we can avoid the anisotropy which is caused by the appearance of magnetic field. The total energy density $\varepsilon$ and total pressure $p$ now write
\begin{eqnarray}
    \varepsilon=\varepsilon_M+\frac{B^2}{8\pi} \label{epsilontotal},\\
    p=p_M+\frac{B^2}{24\pi}\label{ptotal},
\end{eqnarray}
where the subscript $M$ stands for the matter contribution.

\begin{figure}
    \centering
    \includegraphics[width=0.75\linewidth]{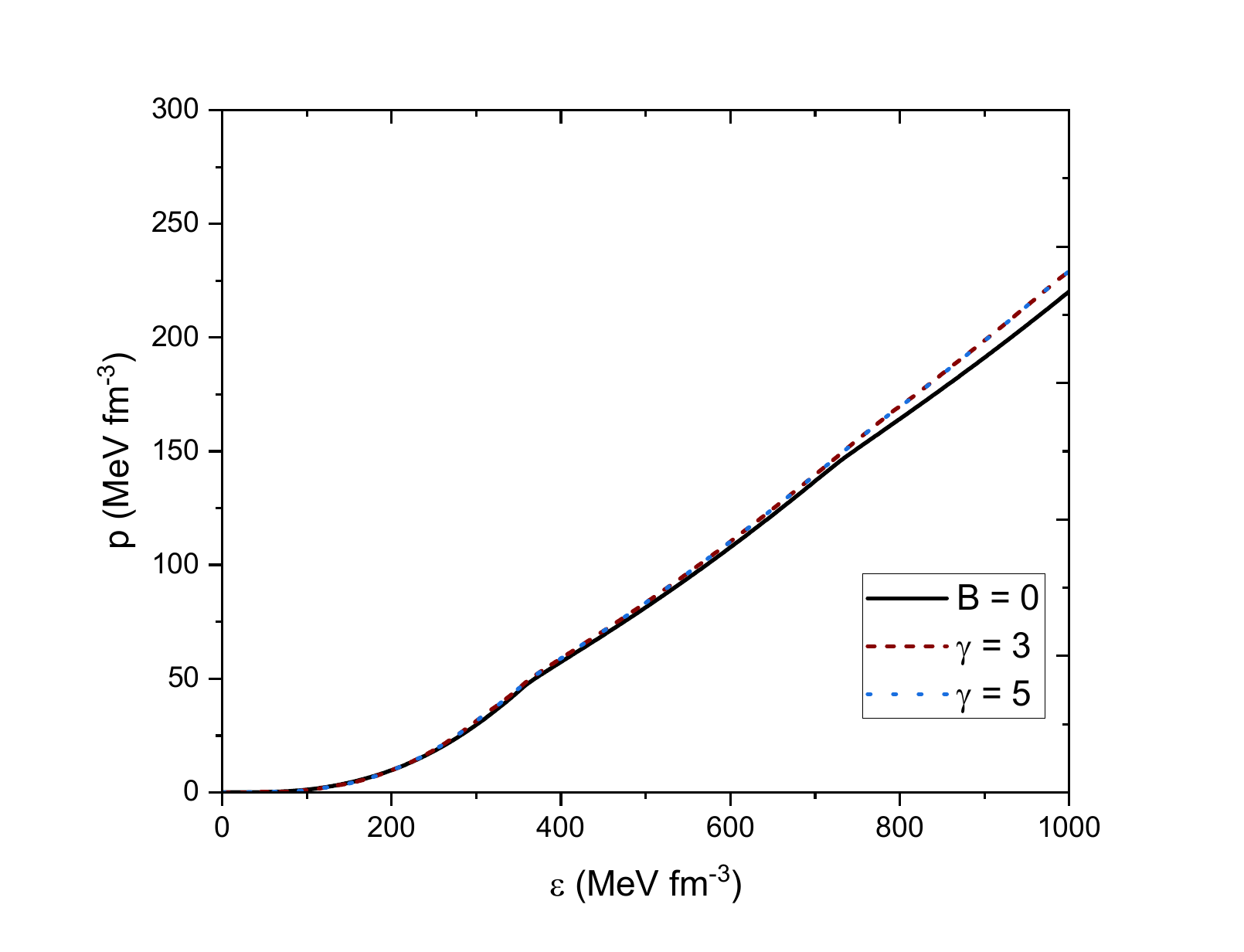}
    \caption{Relation between total energy density $\varepsilon$ and total pressure $p$.}
    \label{eos}
\end{figure}

An ansatz of the magnetic field is proposed in Ref.~[\citen{Lopes2015}], i.e.
\begin{equation}
    B=B_0\left(\frac{\varepsilon_M}{\varepsilon_0}\right)^{\gamma}+B_{surf}.
\end{equation}
Here $B_0$ can be interpreted as the expected magnetic field at the center of the star, $\varepsilon_0$ is the energy density at the center of the NS with maximum mass when the magnetic field is zero in static configuration within GR framework, $\gamma$ is any positive number, and $B_{surf}$ is the magnetic field at the surface of the NSs.

Note, however, that the conventional treatment of magnetic fields in NSs, which produces anisotropy—as presented in Refs.~[\citen{Rizaldy2018b,Rizaldy2018a,Mallick2014}]—faces a major issue: the thermodynamic concept of pressure cannot depend on direction, as it is a scalar quantity~\cite{Lopes2020}. By introducing the concept of a chaotic magnetic field, with the pressure given by $p_B = \varepsilon_B / 3$, where the subscript $B$ stands for the magnetic field, one obtains a pressure that is independent of direction and the choice of coordinate system, thereby restoring the proper thermodynamic concept of pressure and ensuring consistency with field theory~\cite{Lopes2015}.

In Ref.~[\citen{Lopes2015}], it is shown that the Lopes-Menezes chaotic magnetic field ansatz behaves as parameter free when $\gamma \geq 2$. In Ref.~[\citen{Lopes2020}], $\gamma$ is varied to 4 and 6. In this study, we vary $\gamma$ to 3 and 5. Figure~\ref{eos} shows the relation between the total energy density and total pressure, both without and with the magnetic field.

\subsection{Rastall Gravity}
In GR, the covariant divergence of EMT vanishes, i.e. $\nabla_\mu T^{\mu\nu}=0$, while in RG, it writes \cite{Sak} 
\begin{equation}
\nabla_\mu T^{\mu\nu}=\lambda \nabla^{\nu}\mathscr{R}.
\label{covdivRastall}
\end{equation}
Here $T^{\mu\nu}$ represents EMT, $\lambda$ is Rastall parameter that determines the deviation of RG from GR, and $\mathscr{R}$ denotes Ricci scalar. Eq.~(\ref{covdivRastall}) generates a new field equation that is different from Einstein field equation. The Rastall field equation reads
\begin{equation}
G^\mu_\nu+\kappa\lambda \mathscr{R} \delta^\mu_\nu = \kappa T^\mu_\nu,
\label{fieldeq}
\end{equation}
where $G^\mu_\nu$ is the Eisntein tensor, and $\kappa$ is the proportional constant of the field equation. From Eq.~(\ref{fieldeq}), for 4-dimensional spacetime, one can obtain
\begin{equation}
\mathscr{R}=\frac{\kappa T}{4\kappa\lambda - 1},
\label{RicciRastall}
\end{equation}
where $T$ denotes the energy-momentum scalar.

To simplify the Rastall field equation, we can write Eq.~(\ref{fieldeq}) as
\begin{equation}
G^\mu_\nu=\kappa \tilde{T}^\mu_\nu,
\label{fieldeqRastall}
\end{equation}
where
\begin{equation}
\tilde{T}^\mu_\nu=T^\mu_\nu-\delta^\mu_\nu\left(\frac{\kappa\lambda}{4\kappa\lambda-1}T\right).
\label{Teffective}
\end{equation}

Note that in GR, $\kappa = 8\pi$. This value is obtained by applying GR to the weak-field limit. By applying a similar procedure to RG, da Silva et al.~\cite{daSilva2021} approximate $\kappa$ as 
\begin{equation}
\kappa=\frac{8\pi}{2\Lambda+1},
\label{kappa}
\end{equation}
where $\Lambda$ is given by
\begin{equation}
    \lambda=\frac{\Lambda}{\kappa(4\Lambda-1)}
    \label{capitallambda}
\end{equation}

Another important insight for RG comes from Ref.~[\citen{Moradpour2016}], which provides additional information regarding the allowed values of the Rastall parameter. The $\kappa$ must satisfy
\begin{equation}
    {\kappa=\frac{4\kappa\lambda-1}{6\kappa\lambda-1}8\pi,}\label{kappamoradpour}
\end{equation}
and the $\lambda$ must satisfy
\begin{equation}
    {\lambda=\frac{\kappa\lambda(6\kappa\lambda-1)}{(4\kappa\lambda-1)8\pi}.}\label{lambdamoradpour}
\end{equation}

From Eq.~(\ref{RicciRastall}), we see that $\kappa \lambda = 1/4$ leads to a singularity and is therefore forbidden. Similarly, from Eqs.~(\ref{kappa}) and (\ref{capitallambda}), we find that $\Lambda = -1/2$ and $\Lambda = 1/4$ are also not allowed. Moreover, Eq.~(\ref{kappamoradpour}) conveys the same constraint as Eq.~(\ref{RicciRastall}), while Eq.~(\ref{lambdamoradpour}) shows that the condition $\kappa \lambda = 1/6$ is likewise excluded. In this work we use $\lambda=-1\times10^{-5}$, $\lambda=-0.6\times10^{-5}$, $\lambda=0$ (GR), $\lambda=0.6\times10^{-5}$, and $\lambda=1\times10^{-5}$. As expected, Rastall gravity reduces to GR in the limit $\lambda = 0$.
\subsection{Hartle-Thorne Formalism for Rotating Relativistic Stars}

Firstly, it has to be noted that based on Eq. (\ref{epsilontotal}) and Eq. (\ref{ptotal}), it is obvious that the total energy density and total pressure arising from the matter and the presence of a magnetic field can be redefined by introducing $\varepsilon$ and $p$. Therefore, in this subsection, the definitions of $\varepsilon$ and $p$ also encompass the total contributions from both matter and the magnetic field. Consequently, the formulation of HT presented in this subsection is indirectly modified by the presence of the magnetic field. Although this modification is not explicitly apparent, the contribution of the magnetic field is embedded within $\varepsilon$ and $p$.

It is also worth noting that the HT formalism is based on a multipole expansion, in which the monopole and quadrupole terms are typically considered. The HT formalism within RG has been derived in Ref.~[\citen{Pattersons2024}]; however, the derivation only addresses the monopole sector, which corresponds to the mass correction. In this work, we complete the formalism by deriving the quadrupole sector within RG, which accounts for stellar deformation.

In HT formalism, the metric reads \cite{Pattersons2021,Hartle1967,Hartle1968,Pattersons2024}
\begin{eqnarray}
ds^2&=&-e^{2\nu} dt^2+e^{2\lambda}dr^2+r^2\sin^2\theta\:e^{2\psi}(d\phi-\omega\:dt)^2+r^2e^{2\mu}d\theta^2,\label{metricrotstar}
\end{eqnarray}
where $\omega$ is the angular velocity of the local inertial frame, which is proportional to the star's angular velocity $\Omega$ relative to a distant observer. In this case, $\omega$ and $\Omega$ satisfy $\omega=\Omega-\bar{\omega}$, where $\bar{\omega}$ is the angular velocity of the star relative to the local inertial frame.

Due to rotational perturbation, the exponential functions in Eq. (\ref{metricrotstar}) are expanded as the following:
\begin{eqnarray}
e^{2\nu}&=&e^{2\varphi}[1+2(h_0+h_2P_2(\cos\theta))],\\
e^{2\lambda}&=&\left[1+\frac{2}{r}\left(m_0+m_2P_2(\cos\theta)\right)\left(1-\frac{2m(r)}{r}\right)^{-1}\right]\left(1-\frac{2m(r)}{r}\right)^{-1}\\
e^{2\psi}&=&[1+2(v_2-h_2)P_2(\cos\theta)],\\
e^{2\mu}&=&[1+2(v_2-h_2)P_2(\cos\theta)].\label{exponentialfactors}
\end{eqnarray}
Here $h_0$, $h_2$, $m_0$, $m_2$, and $v_2$ are functions of perturbation due to rotation; $P_2(\cos\theta)$ is the second order of Legendre polynomial; $e^{2\varphi}$ is a function which is, in RG, constrained by
\begin{eqnarray}
\frac{d\varphi}{dr}=\frac{m+\frac{\kappa}{2}\tilde{p}}{r(r-2m)},\label{metricPhiRastall}
\end{eqnarray}
while the mass $m$ is constrained by
\begin{eqnarray}
\frac{dm}{dr}=\frac{\kappa}{2} r^2 \tilde{\varepsilon}.
\label{mass}
\end{eqnarray}

In RG, the modified Tolman-Oppenheimer-Volkoff (TOV) equation which represent the hydrostatic equilibrium of the relativistic bodies writes
\begin{eqnarray}
\frac{dp}{dr}&=&\left[-\frac{\left(\varepsilon+p\right)\left(m+\frac{\kappa}{2}r^3 \tilde{p}\right)}{r(r-2m)}\right]\left[1-\frac{\kappa\lambda}{4\kappa\lambda-1}\left(3-\frac{d\varepsilon}{dp}\right)\right]^{-1}.
\label{TOVR}
\end{eqnarray}

For a rotating configuration, $\bar{\omega}$ can be calculated by solving the following equation \cite{Hartle1967}:
\begin{equation}
    \frac{1}{r^4}\frac{d}{dr}\left(r^4 j\frac{d\bar{\omega}}{dr}\right)+\frac{4}{r}\frac{dj}{dr}=0,
\label{omega}
\end{equation}
where
\begin{equation}
    j=e^{-\varphi}\left(1-\frac{2m}{r}\right)^{1/2}.
\end{equation}
Note that the explicit expression in Eq.~(\ref{omega}) is identical in both GR and RG~\cite{Pattersons2024}. The boundary condition at $r=0$ is $\bar{\omega}=\omega_c$, where $\omega_c$ can be chosen arbitrarily.

It is important to note that Eqs.~(\ref{metricPhiRastall}), (\ref{mass}), and (\ref{TOVR}) describe the static configuration. However, in the presence of rotation, corrections to the mass and radius must be formulated. To obtain these corrections, the Rastall field equations must be applied to the perturbative expansion terms.

We followed HT procedure \cite{Hartle1967} to calculate the corrections of Einstein tensor to obtain the RG's version of the corrections of Einstein tensor, i.e.
\begin{eqnarray}
    \Delta G^\mu_\nu=\delta G^\mu_\nu - \kappa\xi\frac{d}{dr}\tilde{T}^\mu_\nu.
\end{eqnarray}
Here, $\delta G^\mu_\nu$ denotes the expansion terms of $G^\mu_\nu$. As presented in the original paper of HT formalism~\cite{Hartle1967}, and also adopted in Ref.~[\citen{Pattersons2021}], we limit the expansion terms of $\xi$ to the $l=2$, i.e.
\begin{equation}
\xi=\xi_0(r)+\xi_2(r) P_2(\cos\theta).
\end{equation}
The relations between $\xi$ and pressure perturbation factor $\mathscr{P}$ writes
\begin{eqnarray}
\xi_0=-\mathscr{P}_0\left(\varepsilon+p\right)\left(\frac{dp}{dr}\right)^{-1}, \label{xi0}\\
\xi_2=-\mathscr{P}_2\left(\varepsilon+p\right)\left(\frac{dp}{dr}\right)^{-1}. \label{xi2}
\end{eqnarray}

The corrections of energy-momentum scalar reads
\begin{eqnarray}
\left(\Delta T\right)_{l=0}&=&-\frac{2}{3}\left(\varepsilon+p\right)e^{-2\varphi}r^2\bar{\omega},\\
\left(\Delta T\right)_{l=2}&=&-2\left(\varepsilon+p\right)\frac{d\varepsilon}{dp}.
\end{eqnarray}

The corrections of the Einstein tensor now take the form as follows, i.e.
\begin{equation}
\Delta G^\mu_\nu=\kappa\Delta\tilde{T}^\mu_\nu=\kappa\left[\Delta T^\mu_\nu-\delta^\mu_\nu\left(\frac{\kappa\lambda}{4\kappa\lambda-1}\Delta T\right)\right].
\end{equation}

By calculating the Rastall field equation, for the ($tt$)-component of the $l=0$ order, we can obtain
\begin{eqnarray}
\frac{dm_0}{dr}&=&\frac{\kappa}{2}r^2 \frac{d\tilde{\varepsilon}}{dp}\left(\varepsilon+p\right)\mathscr{P}_0\nonumber+\frac{1}{12}j^2r^4\left(\frac{d\bar{\omega}}{dr}\right)^2\\
&&-\frac{1}{3}r^3\left(\frac{dj^2}{dr}\right)\bar{\omega}^2 \left(1-\frac{\kappa\lambda}{4\kappa\lambda-1}\right).
\label{m0}
\end{eqnarray}
From the ($rr$)-component of the $l=0$ order, we can obtain
\begin{eqnarray}
\frac{d\mathscr{P}_0}{dr}&=&-\frac{m_0\left(1+\kappa r^2 \tilde{p}\right)}{\left(r-2m\right)^2}-\frac{\kappa r^2\left(\varepsilon+p\right)}{2\left(r-2m\right)}\left[1+\frac{\kappa\lambda}{4\kappa\lambda-1}\left(1-\frac{d\varepsilon}{dp}\right)+\frac{2\kappa\lambda}{4\kappa\lambda-1}\right]\:\mathscr{P}_0\nonumber\\
&&+\frac{1}{3}\left[\frac{d}{dr}\left(\frac{r^3j^2\bar{\omega}^2}{r-2m}\right)-\left\{\frac{r^3}{r-2m}\frac{\kappa\lambda}{4\kappa\lambda-1}\left(\frac{dj^2}{dr}\right)\bar{\omega}^2\right\}\right]\nonumber\\
&&+\frac{1}{12}\frac{r^4j^2}{r-2m}\left(\frac{d\bar{\omega}}{dr}\right)^2.
\label{p_0}
\end{eqnarray} 
Here $\mathscr{P}_0$ denotes the pressure perturbation factor. The boundary conditions are $m_0(r=0)=\mathscr{P}_0(r=0)=0$.

Mass correction $\delta M$ of the star is given by
\begin{eqnarray}
    \delta M = m_0(R)+\frac{L^2}{R^3},
\end{eqnarray}
where $R$ denotes the radius of the star, and $L$ denotes the angular momentum, which satisfies
\begin{eqnarray}
    L=\frac{\kappa}{3}\int^R_0 dr\:r^4\frac{\varepsilon+p}{\left(1-\frac{2m}{r}\right)^{1/2}}\:\bar{\omega}\:e^{-\varphi}.
    \label{angularmomentum}
\end{eqnarray}
Thus, the total mass reads
\begin{equation}
    M=M_0+\delta M,
\end{equation}
where $M_0$ is the mass of NSs within static configuration which is obtained by solving Eqs. (\ref{metricPhiRastall}), (\ref{mass}), and (\ref{TOVR}) simultaneously.

Now we proceed to the quadrupole sector which is not considered in Ref.~[\citen{Pattersons2021}]. From the ($\theta\theta$), ($\phi$,$\phi$), ($\theta r$), and ($rr$) components of the $l=2$ order of Einstein field equation, we obtain
\begin{eqnarray}
    \frac{dv_2}{dr}=-2\frac{d\varphi}{dr}h_2+\left(\frac{1}{r}+\frac{d\varphi}{dr}\right)\left[-\frac{1}{3}r^3\frac{dj^2}{dr}\bar{\omega}^2+\frac{1}{6}j^2r^4\left(\frac{d\bar{\omega}}{dr}\right)^2\right],
\end{eqnarray}
\begin{eqnarray}
     \frac{dh_2}{dr}&=&-2\frac{d\varphi}{dr}h_2+\frac{\left[\kappa r (\varepsilon+p)\left(\frac{1+2\kappa^2\lambda}{4\kappa-1}\frac{d\varepsilon}{dp}\right)-\frac{4m}{r^2}\right]}{(r-2m)\left(2\frac{d\varphi}{dr}\right)}h_2-\frac{4v_2}{r(r-2m)\left(2\frac{d\varphi}{dr}\right)}\nonumber\\
     &&+\frac{r^3j^2}{6}\left[\frac{d\varphi}{dr}r-\frac{1}{(r-2m)\left(2\frac{d\varphi}{dr}\right)}\right]\left(\frac{d\bar{\omega}}{dr}\right)^2-\frac{r^2\bar{\omega}^2}{3}\left[\frac{d\varphi}{dr}+\frac{1}{(r-2m)\left(2\frac{d\varphi}{dr}\right)}\right]\nonumber\\
     &&\times \frac{dj^2}{dr}-\frac{2}{3}\frac{\kappa\lambda}{4\kappa\lambda-1}\left(\frac{dj^2}{dr}\right)r\bar{\omega}^2\left[\frac{r}{(r-2m)\left(2\frac{d\varphi}{dr}\right)}\right]\frac{d\varepsilon}{dp}.
\end{eqnarray}
The boundary conditions at $r=0$ are $v_2(0)=0$ and $h_2=0$.

There exists a formula \cite{Pattersons2021,Hartle1967,Occhionero1969}
\begin{equation}
    \mathscr{P}_2=-h_2-\frac{1}{3}r^2e^{-2\varphi}\bar{\omega}^2.
    \label{formula}
\end{equation}
Eq. (\ref{formula}) shows that by obtaining $h_2$ we can calculate $\mathscr{P}_2$.

The radius correction of the star is given by
\begin{equation}
    \delta r= \xi_0(R)+\xi_2(R) P_2(\cos \theta),
\end{equation}
where
\begin{eqnarray}
    \xi_0=-\mathscr{P}_0(\epsilon+p)\left(\frac{dp}{dr}\right)^{-1},\nonumber\\
    \xi_2=-\mathscr{P}_2(\epsilon+p)\left(\frac{dp}{dr}\right)^{-1}.
\end{eqnarray}
Finally, we can calculate the radius of the pole $R_{POL}$ and the radius of the equator $R_{EQ}$ \cite{Pattersons2021,Lopes2024}
\begin{eqnarray}
    R_{POL}=R+\xi_0+\xi_2, \label{Rpol}\\
    R_{EQ}=R+\xi_0-\frac{\xi_2}{2}. \label{Req}
\end{eqnarray}
This allows us to calculate the eccentricity
\begin{equation}
    e=\sqrt{1-\frac{R_{POL}^2}{R_{EQ}^2}}.
\end{equation}
Note that by setting $\lambda=0$, the formulations recover to the standard HT formalism in GR.


Moreover, another important physical quantity to consider is the moment of inertia. With the angular momentum angular $L$ and the velocity relative to the distant observers $\Omega$ in our hands, we can determine the moment of inertia \cite{Hartle1967,daSilva2021,Glendenning1998}
\begin{equation}
    I=\frac{L}{\Omega}. \label{momenofinersia}
\end{equation}

\section{Results and Discussion}
We numerically solved all differential equations using Euler method. {For the value of $\Omega$, we consider $\Omega = 1000$ s$^{-1}$ and $\Omega = 2000$ s$^{-1}$. However, except for the calculation of the moment of inertia, we use $\Omega = 50$ s$^{-1}$, as we only employ the very slow rotation approximation for this purpose. For the initial value of $\bar{\omega}$ (i.e., $\omega_c$), we adopt $\omega_c = 80$ s$^{-1}$ in all cases throughout this work, except for the calculation of the moment of inertia, where $\omega_c = 0$. It is worth noting that the value $\omega_c = 80$ used in our calculations is significantly smaller than the value used in Ref.~[\citen{Wen}], where $\bar{\omega} = 3000$ s$^{-1}$.} Moreover, it is also worth noting that $\bar{\omega}(r)$ increases with $r$, which would result in a large $\bar{\omega}$ value at the surface if a large initial $\omega_c$ is chosen. On the other side, we have insight that according to Ref.~[\citen{Ber}], the HT formalism is accurate to better than 1 per cent even for the fastest millisecond pulsars. However, putting $\bar{\omega}_c$ = 3000 s$^{-1}$ in the calculation is still potential to be dangerous, for the HT formalism is basically an approximation for slowly rotating relativistic stars. So our value of $\bar{\omega}_c$ is safer than the one in Ref.~[\citen{Wen}].
\begin{figure}
    \centering
    \includegraphics[width=1.0\linewidth]{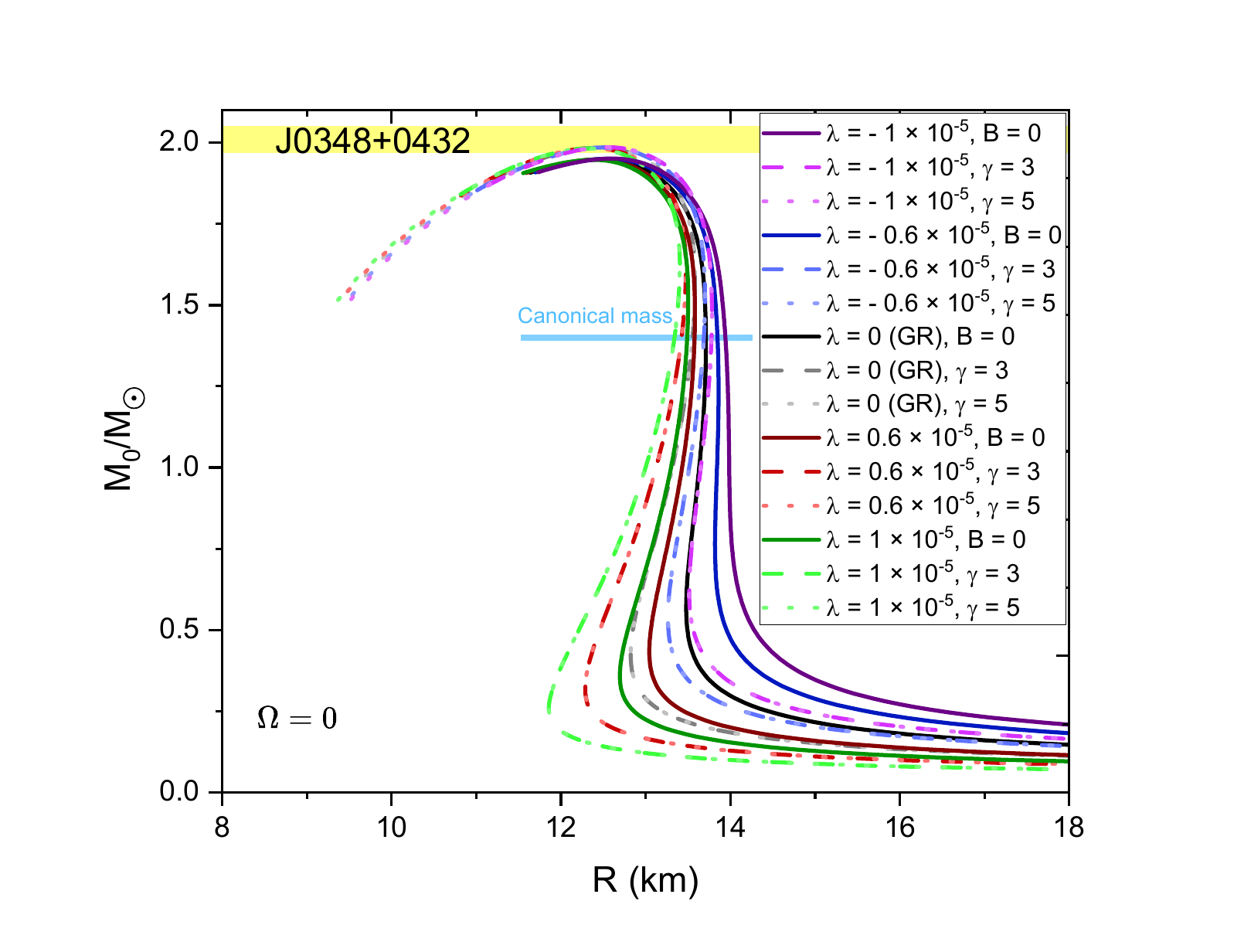}
    \caption{Mass-radius relation of NSs with and without chaotic magnetic field within static configuration.}
    \label{fig:static}
\end{figure}

\begin{figure*}
	\centering
	\resizebox{0.5\textwidth}{!}{\includegraphics{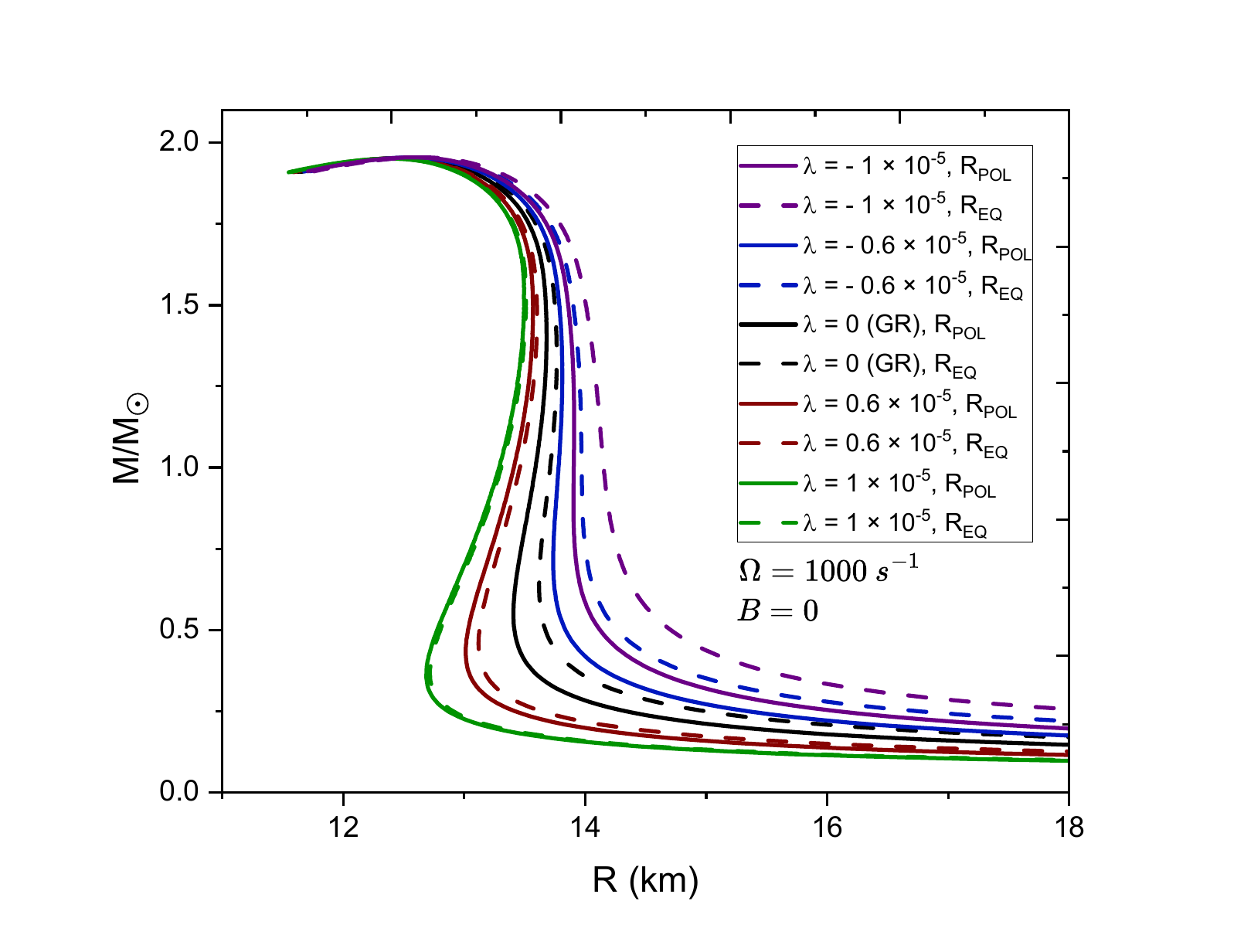}}\resizebox{0.50\textwidth}{!}{\includegraphics{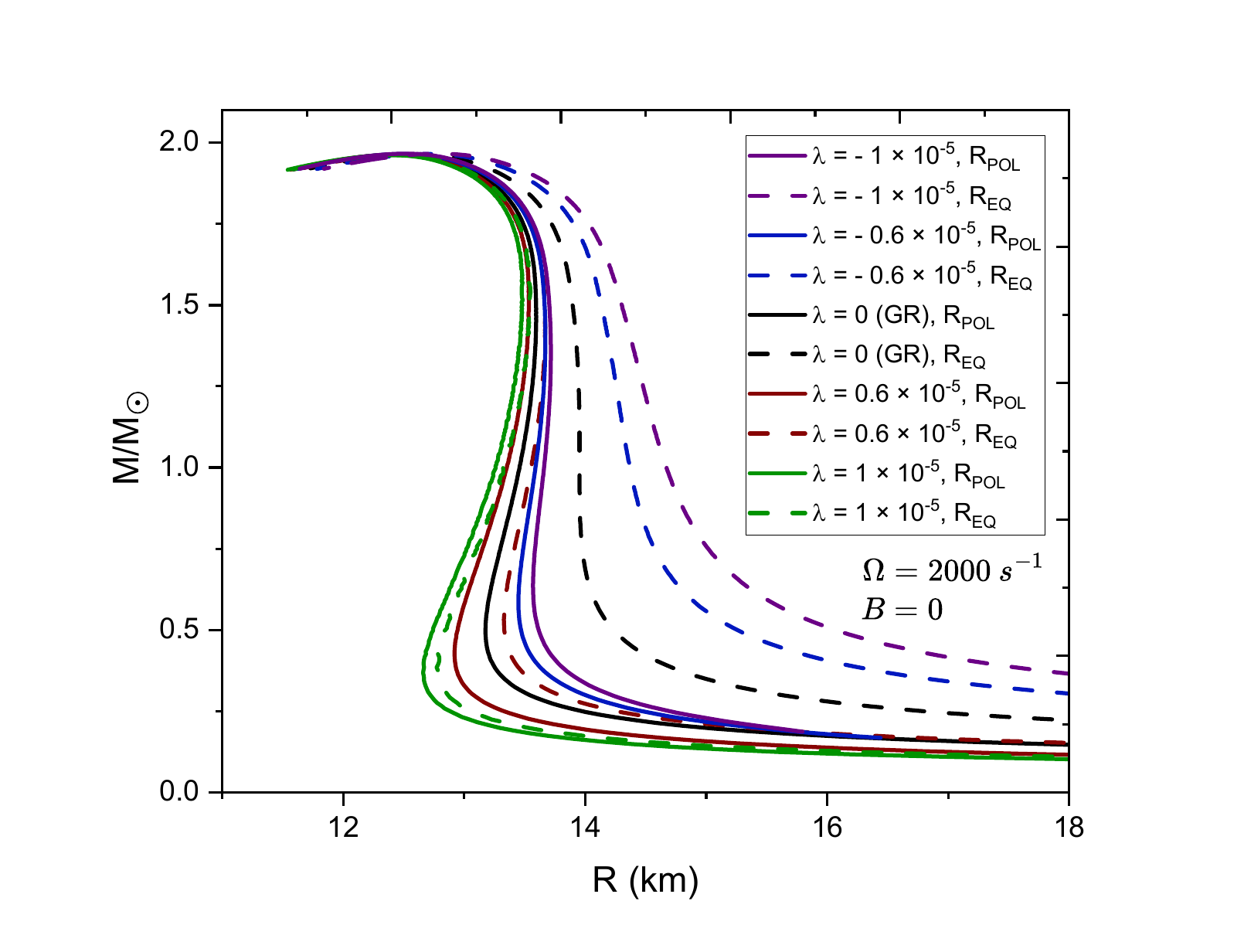}}  \\~~~~~~~~~~~~~~~~~~~~(a)~~~~~~~~~~~~~~~~~~~~~~~~~~~~~~~~~~~~~~~~~~~~~~~~~~~(b)~~~~~~~~~~~~~~~~		\vspace{0.1cm}\\
        \centering
	\resizebox{0.5\textwidth}{!}{\includegraphics{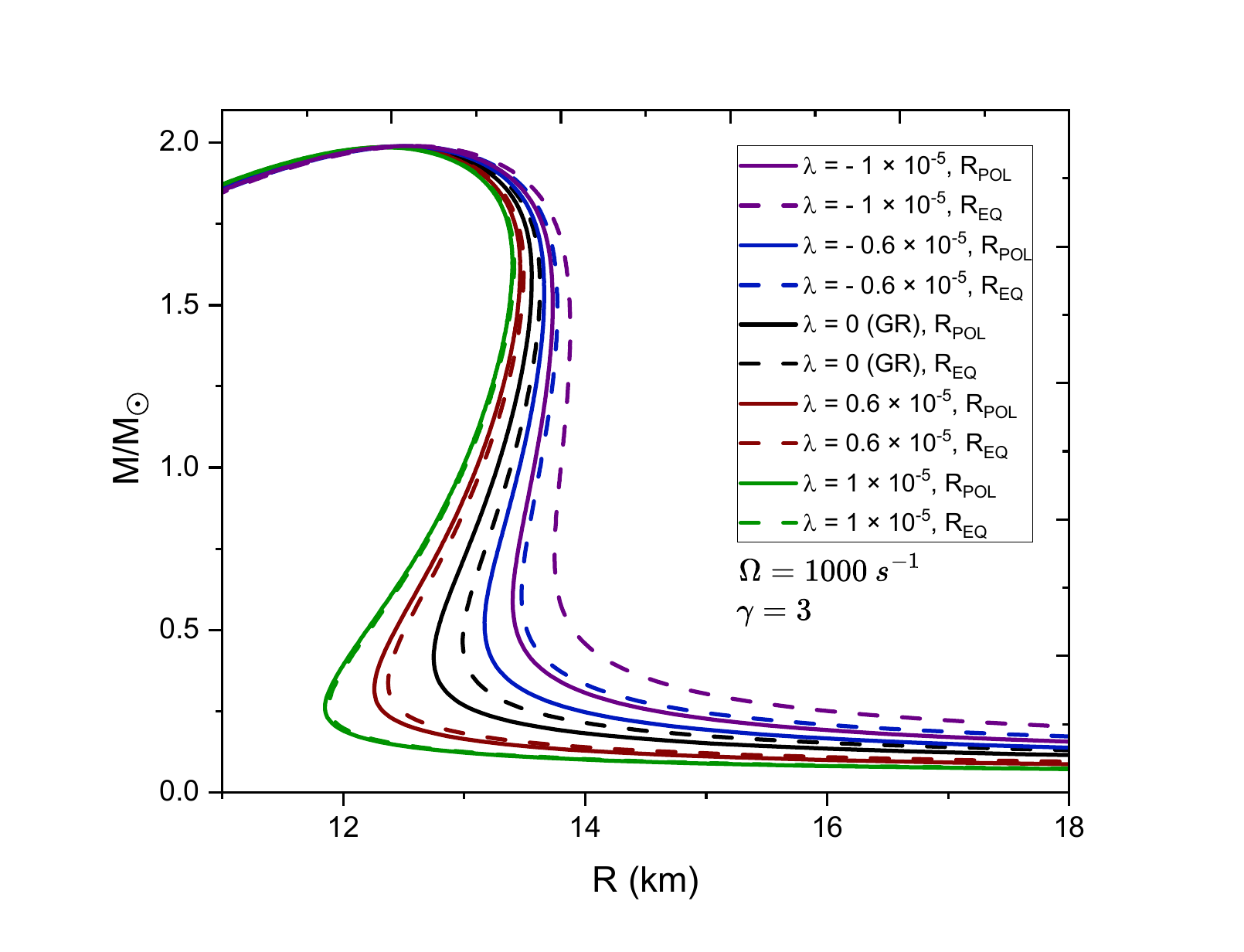}}\resizebox{0.50\textwidth}{!}{\includegraphics{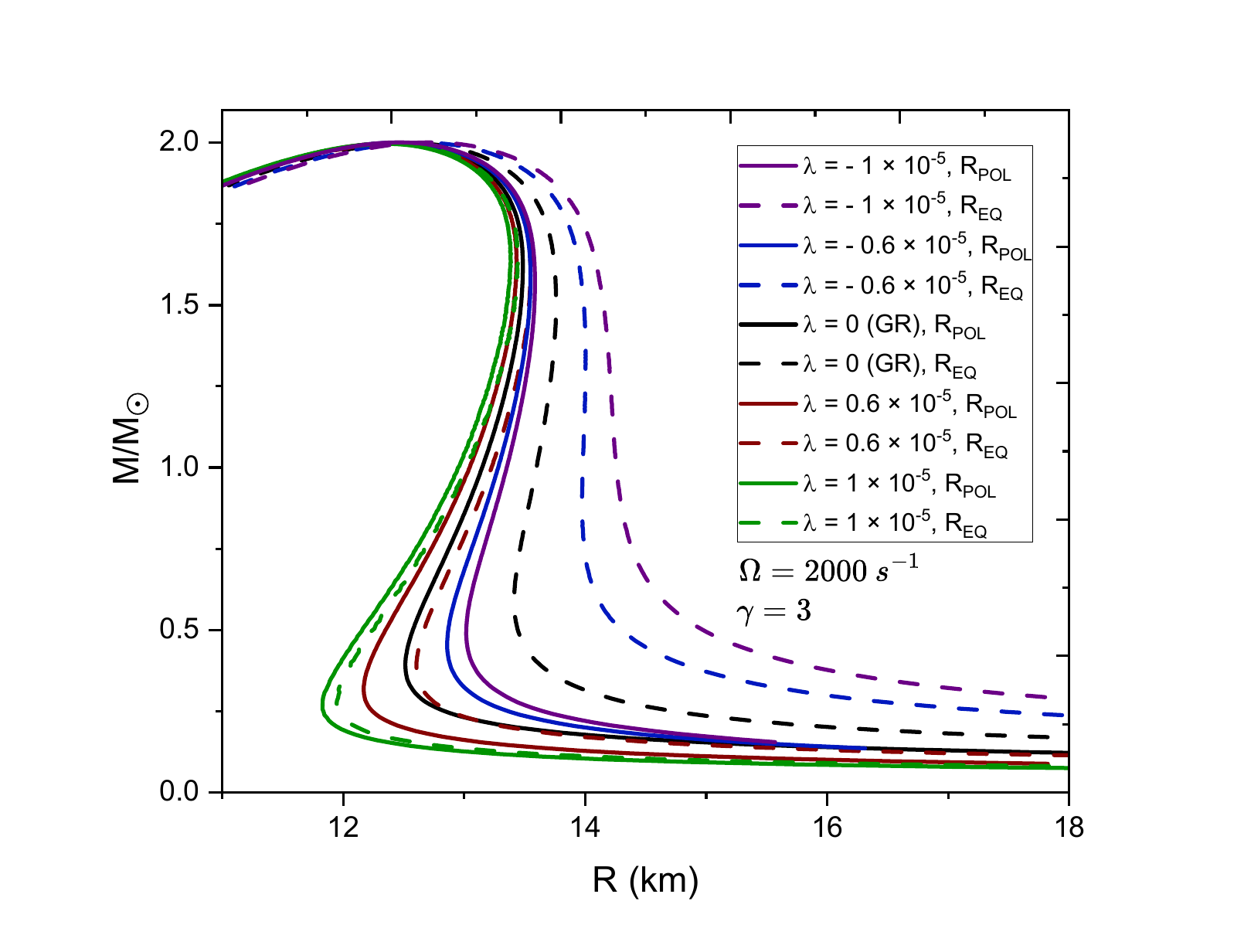}}  \\~~~~~~~~~~~~~~~~~~~~(c)~~~~~~~~~~~~~~~~~~~~~~~~~~~~~~~~~~~~~~~~~~~~~~~~~~~(d)~~~~~~~~~~~~~~~~		\vspace{0.1cm}\\
        \centering
	\resizebox{0.5\textwidth}{!}{\includegraphics{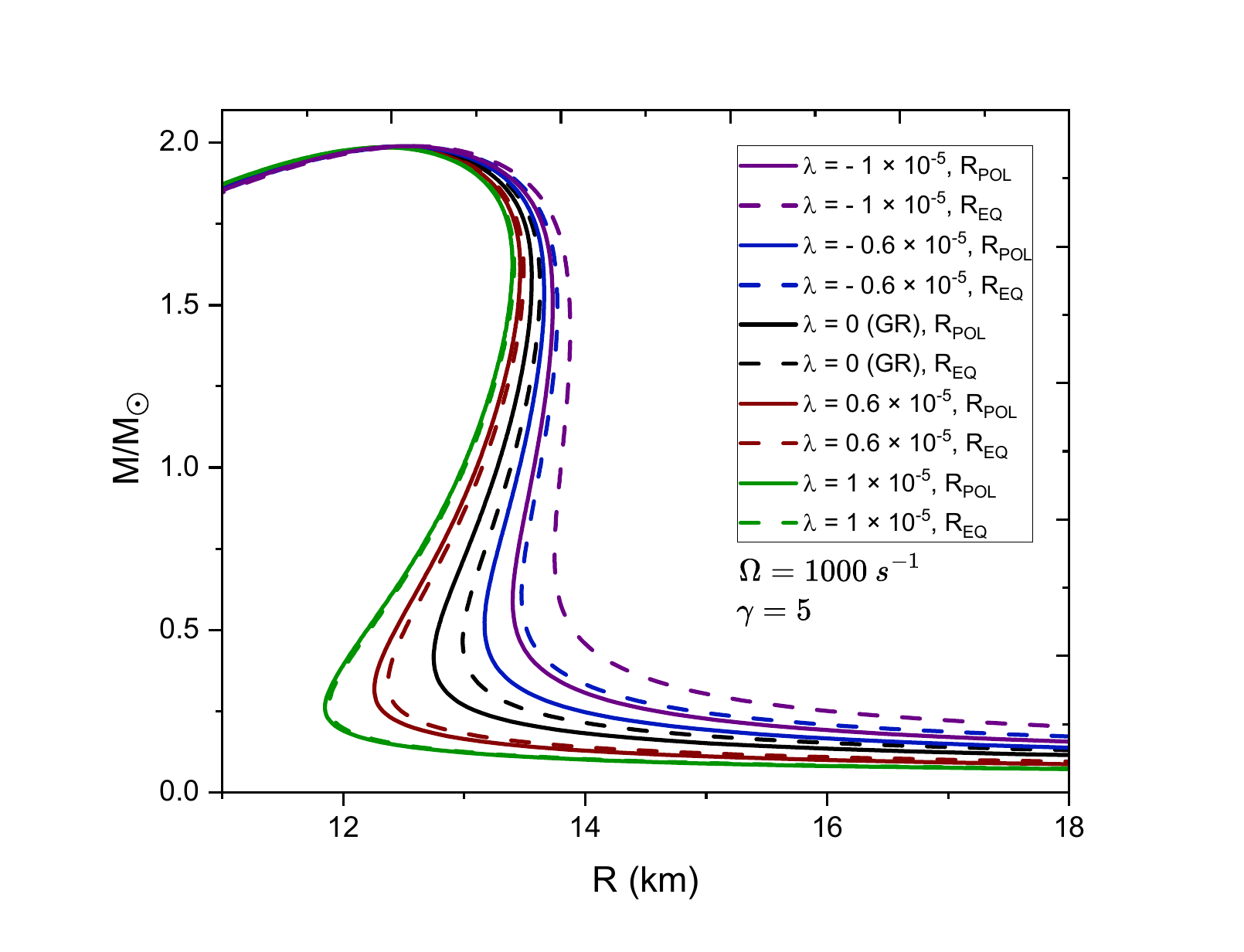}}\resizebox{0.50\textwidth}{!}{\includegraphics{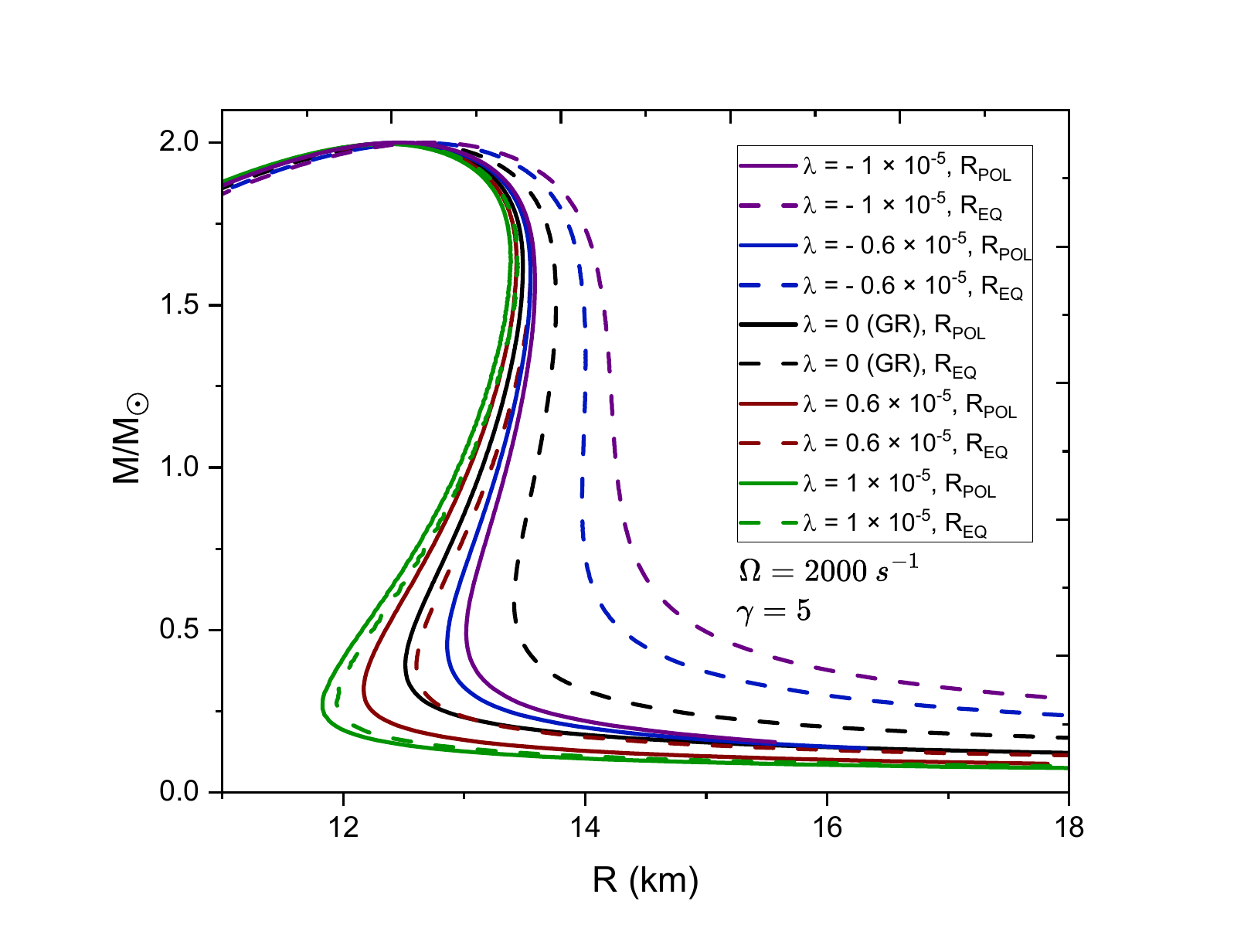}}  \\~~~~~~~~~~~~~~~~~~~~(e)~~~~~~~~~~~~~~~~~~~~~~~~~~~~~~~~~~~~~~~~~~~~~~~~~~~(f)~~~~~~~~~~~~~~~~
	 
	\caption{Mass-radius relation of NSs for (a) $\Omega = 1000$ s$^{-1}$ and $B=0$, (b) $\Omega = 2000$ s$^{-1}$ and $B=0$, (c) $\Omega = 1000$ s$^{-1}$ and $\gamma=3$, (d) $\Omega = 2000$ s$^{-1}$ and $\gamma=3$, (e) $\Omega = 1000$ s$^{-1}$ and $\gamma=5$, and (f) $\Omega = 2000$ s$^{-1}$ and $\gamma=5$.}\label{fig:deform}
 \end{figure*}

\begin{figure*}
	\centering
	\resizebox{0.40\textwidth}{!}{\includegraphics{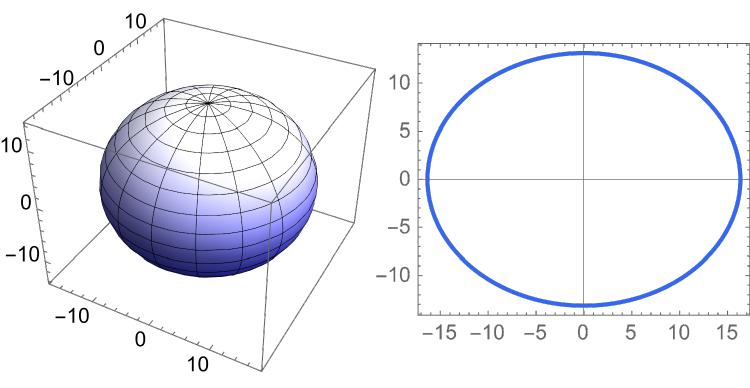}}~~~~~\resizebox{0.40\textwidth}{!}{\includegraphics{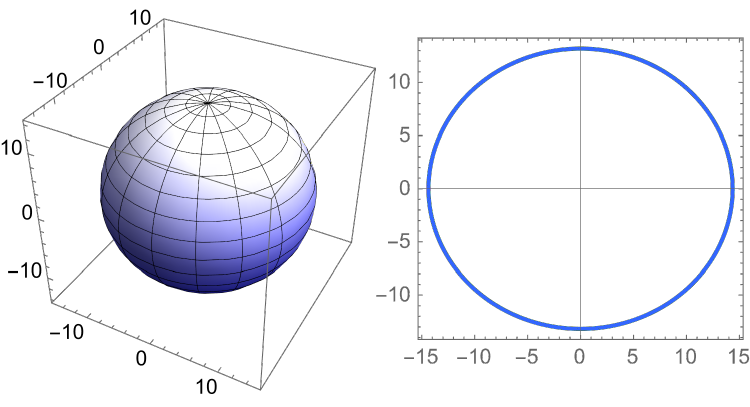}}~~~~~~~~~~~~    \\~~~(a)~~~$ M=0.35\:M_\odot  $~~~~~~~~~~~~~~~~~~~~~~~(b)~~~$ M=0.78\:M_\odot $~~~~~~~~~~~~\\
    \resizebox{0.4\textwidth}{!}{\includegraphics{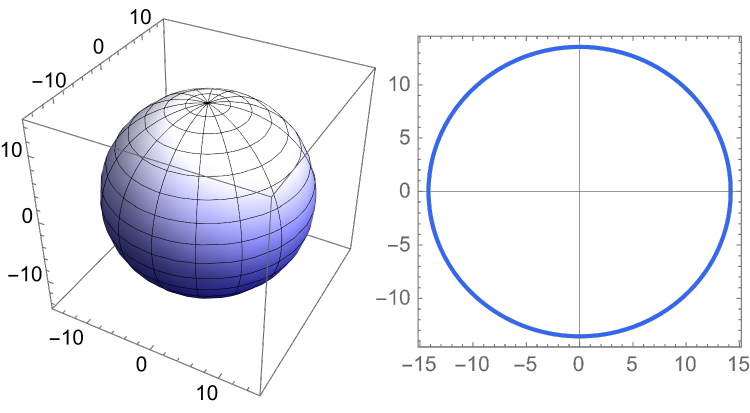}}~~~~~\resizebox{0.4\textwidth}{!}{\includegraphics{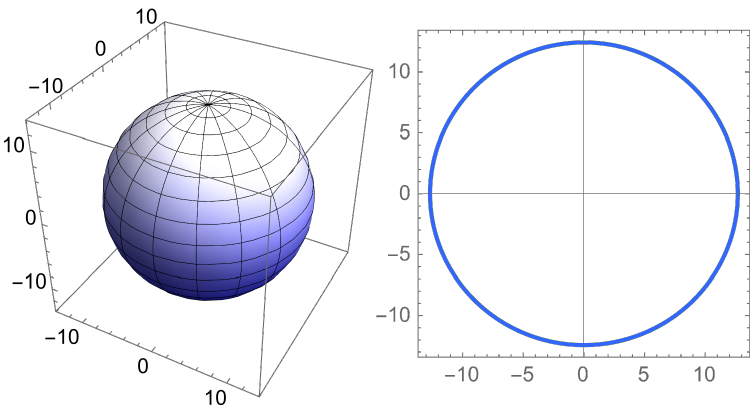}}~~~~~~~~~~ \\~~~~(c)~~~$ M=1.40\:M_\odot $~~~~~~~~~~~~~~~~~~~~~~~(d)~~~$ M=2.00\:M_\odot $~~~~~~~~~~~~ 	
	\caption{Illustrations of 3- and 2-dimensional deformation of NSs in each particular mass at $\lambda=-1\times10^{-5}$, $\gamma=5$, and $\Omega=2000$ s$^{-1}$.}
	\label{fig:deformation}  
\end{figure*}

\begin{figure}
    \centering
    \includegraphics[width=0.75\linewidth]{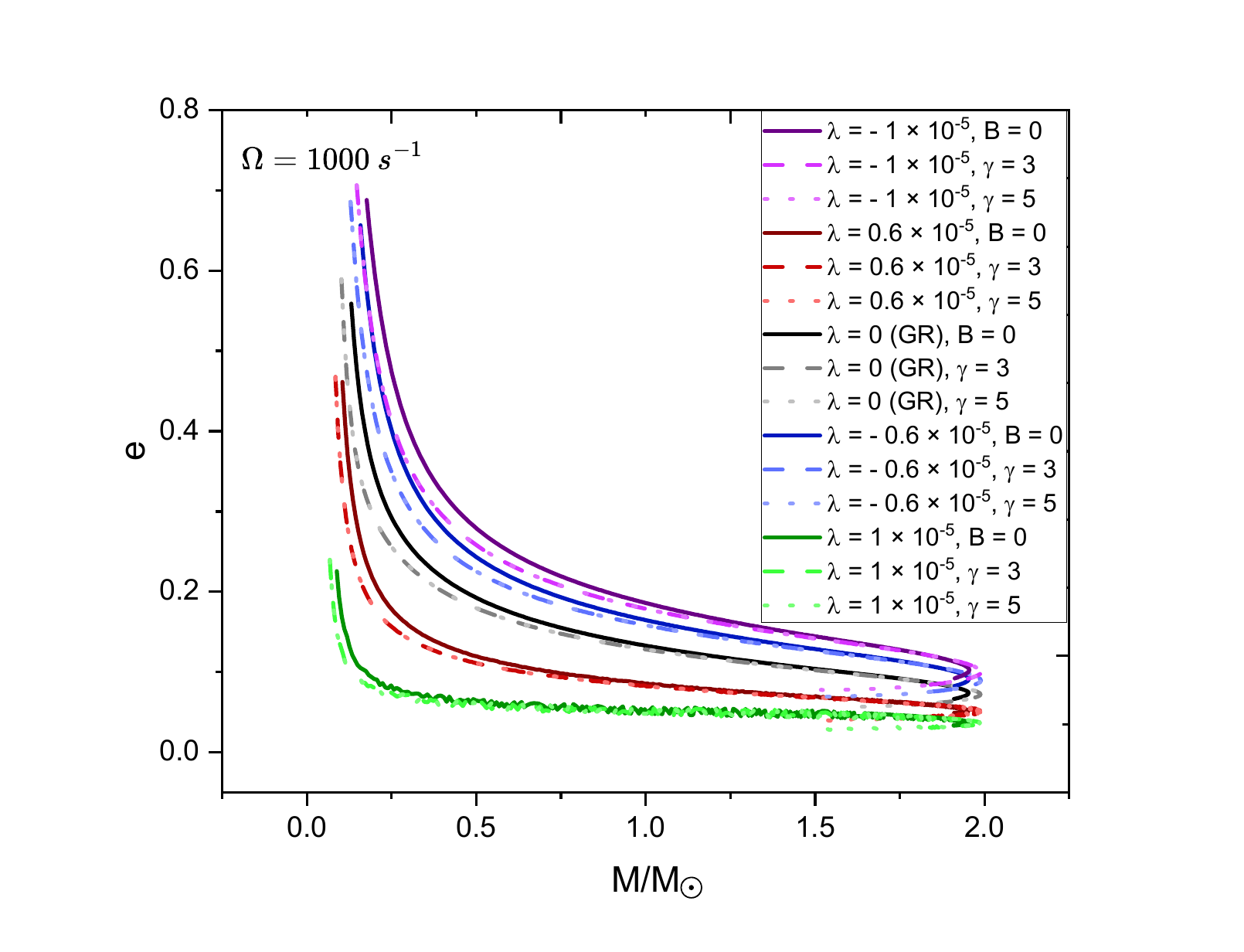}\\
    ~~~~~(a) \vspace{0.1cm}\\
    \centering
    \includegraphics[width=0.75\linewidth]{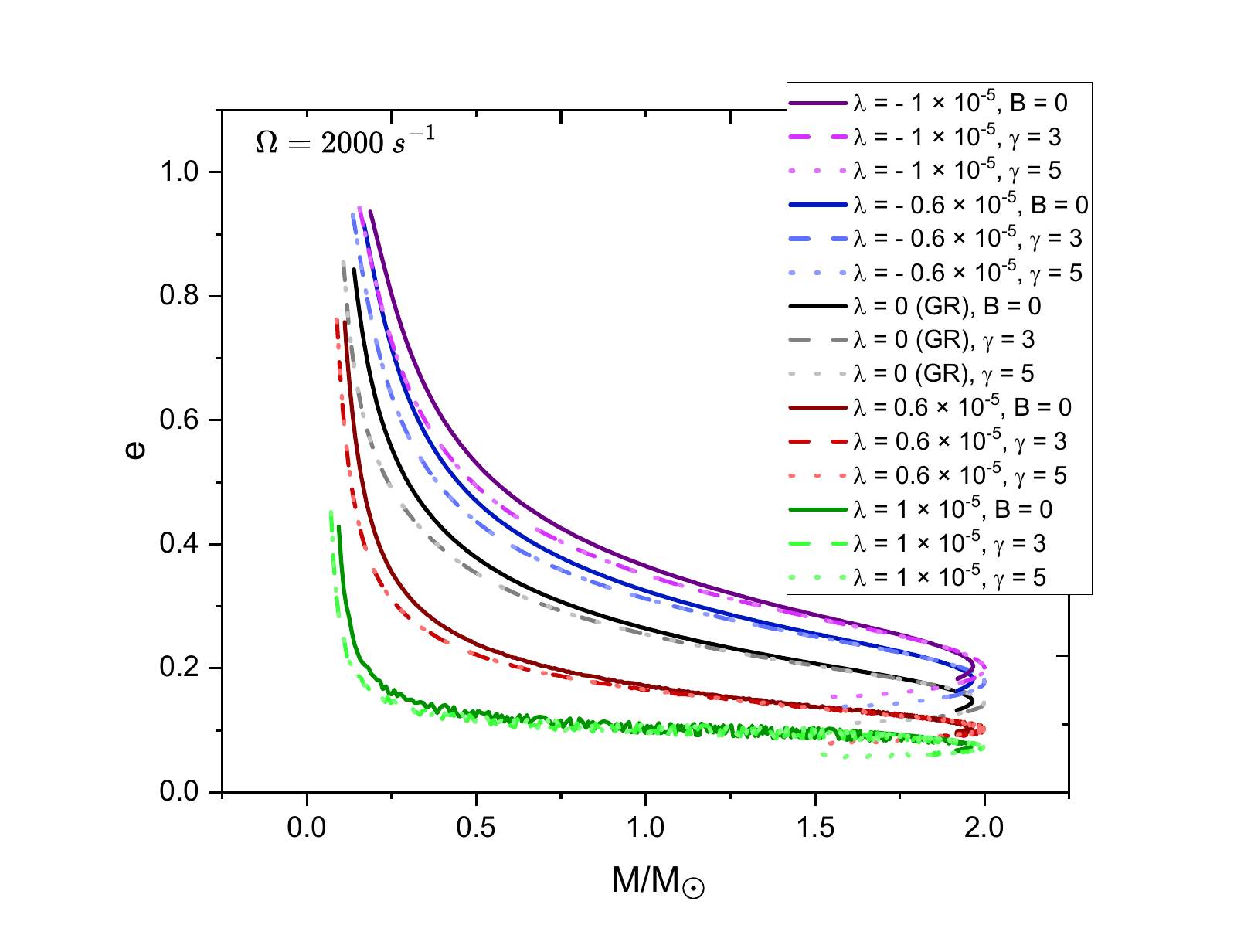}\\
    ~~~~~(b)
    \caption{Eccentricity of NSs at (a) $\Omega =$ 1000 s$^{-1}$, (b) $\Omega =$ 2000 s$^{-1}$ as a function of mass}
    \label{fig:eccentricity}
\end{figure}

\begin{figure}
    \centering
    \includegraphics[width=0.75\linewidth]{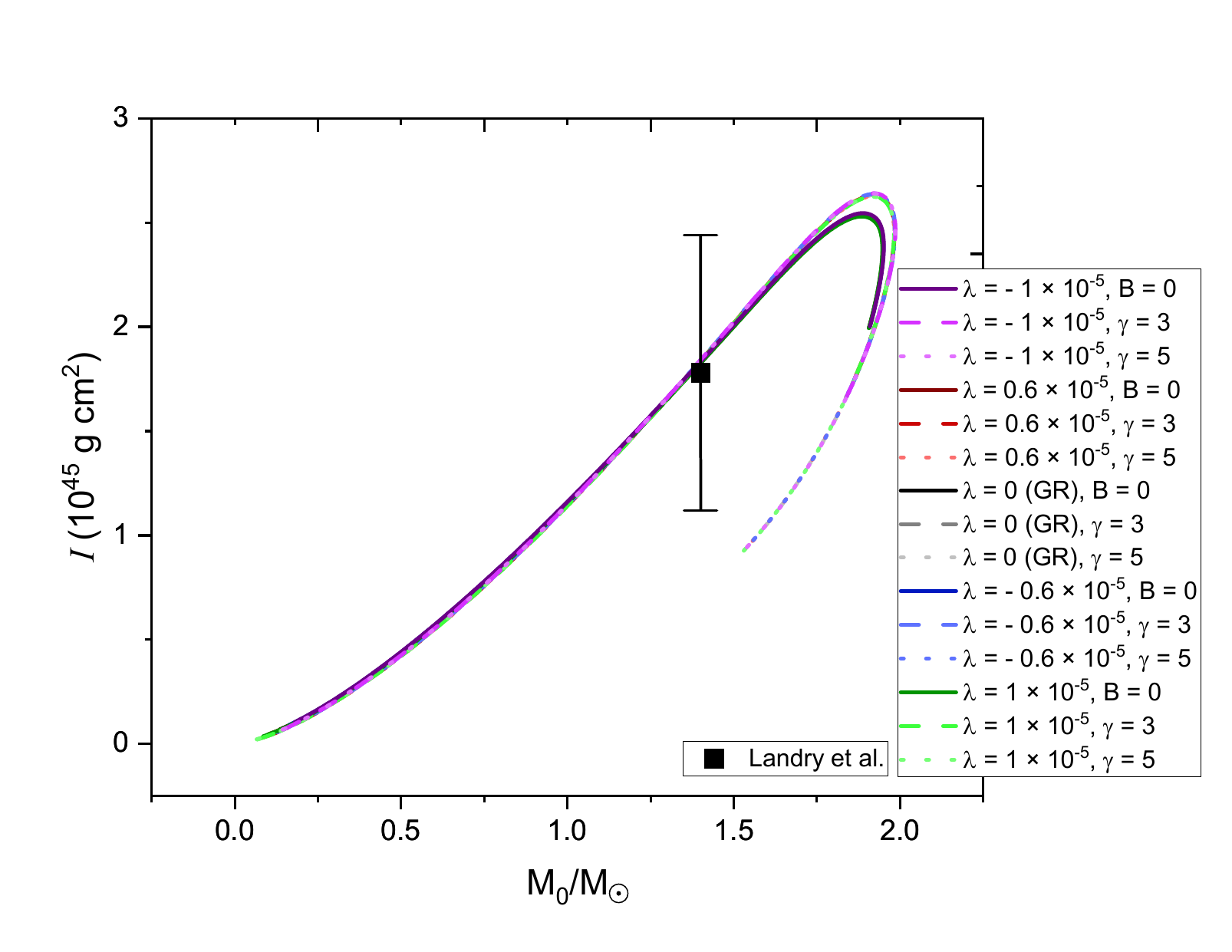}\\
    ~~~~~(a) \vspace{0.1cm}\\
    \centering
    \includegraphics[width=0.75\linewidth]{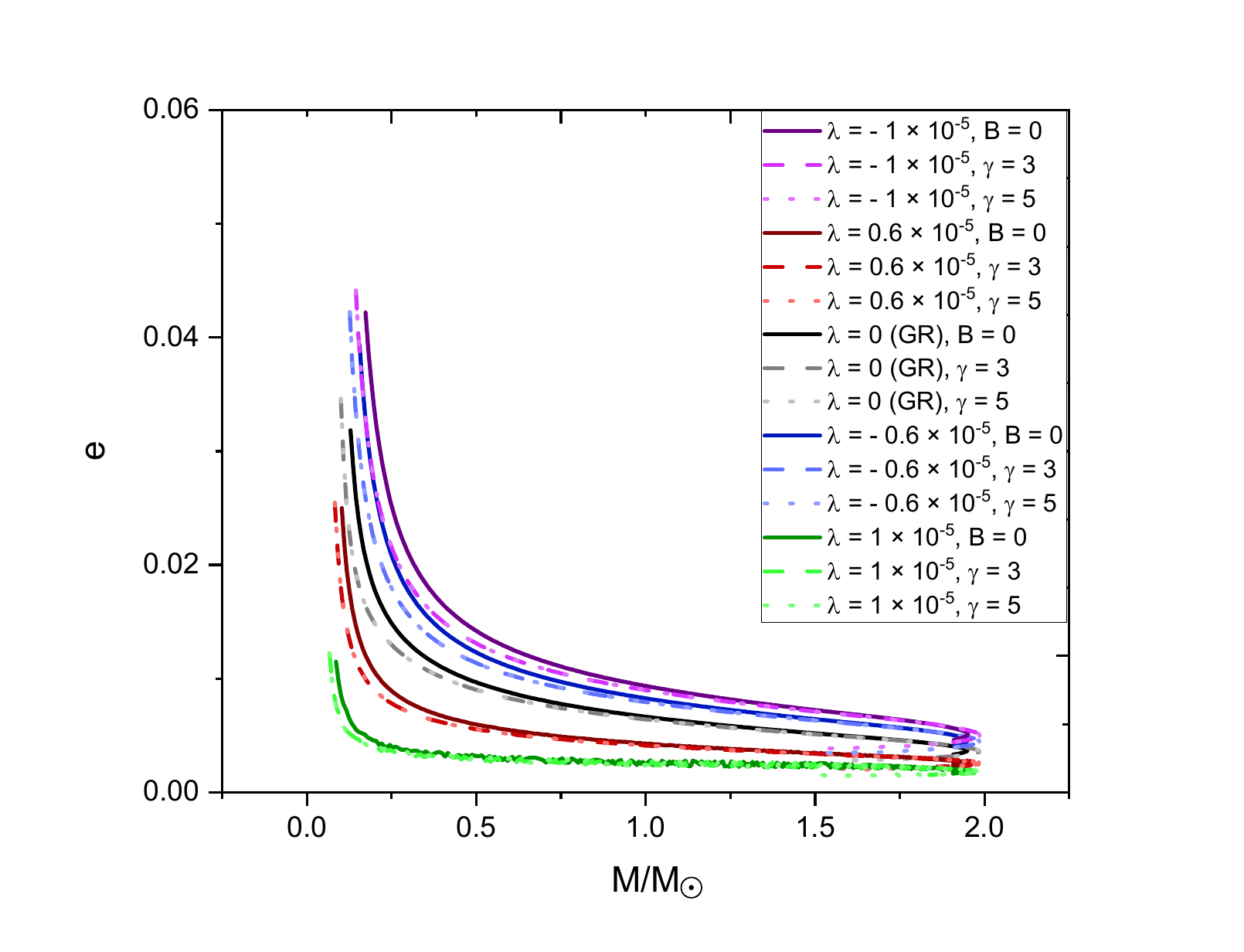}\\
    ~~~~~(b)
    \caption{Moment of inertia of NSs when (a) $\Omega =$ 50 s$^{-1}$, (b) the corresponding eccentricity of the NSs when $\Omega =$ 50 s$^{-1}$. Here we use a constraint range presented by Landry et al.~[\citen{Landry2020}]}
    \label{fig:momin}
\end{figure}

{If we compare our initial value of $\bar{\omega}$—excluding the one used for the moment of inertia calculation—to those in Ref.~[\citen{Pattersons2021}] and Ref.~[\citen{Ordaz2019}], which adopted $\bar{\omega}_c = 0$, our choice of $\bar{\omega}_c = 80$ s$^{-1}$ appears to be more realistic for modeling general rotating NSs. In contrast, their assumption of $\bar{\omega}_c = 0$ is more appropriate only for very slowly rotating neutron stars. A similar choice of $\omega_c$ and the reasoning behind it have also been emphasized by the authors in Ref.~[\citen{Pattersons2024}]. We set the limitation of $\Omega$ to be 2000 s$^{-1}$ in our analysis.}

{We consider both scenarios: without and with the presence of a magnetic field. We employ $B_0=3\times10^{18}$ G and $B_{surf}=1\times10^{15}$ G. As mentioned in Subsection 2.2, we vary the parameter $\gamma$ with values of 3 and 5. It is also important to note that although $\gamma$ is formally a free parameter and can, in principle, take any positive value. However, choosing a very small value (e.g., of the order of 10$^{-3}$) is problematic. In such cases, the magnetic field becomes strongly dominated by the constant component $B_0$, making it effectively resemble a uniform magnetic field, which is unphysical in the context of our model.}

As mentioned in Subsection~2.3, we adopt the values $\lambda = -1 \times 10^{-5}$, $\lambda = -0.6 \times 10^{-5}$, $\lambda = 0$ (GR), $\lambda = 0.6 \times 10^{-5}$, and $\lambda = 1 \times 10^{-5}$. These values are smaller than those used in Ref.~[\citen{Pattersons2024}]; however, they still have a significant impact on the results---greater, in fact, than the impact of the Rastall parameter reported in Ref.~[\citen{Pattersons2024}]. This difference may be attributed to the different EOS employed in this work, which is stiffer than that used in the previous study. This suggests that the influence of the Rastall parameter becomes more pronounced when a stiffer EOS is considered. Nevertheless, the use of other EOSs that have a more significant impact on increasing the maximum mass can be left for future work.

In Appendix~\ref{app:macroprops}, we provide examples of how changes in $\gamma$ affect several key quantities, namely the maximum mass $M^{\text{max}}$, $\langle R \rangle^{\text{max}}$, and $\langle R \rangle^{(1.4)}$. We adopt the definition of $\langle R \rangle$ from Ref.~[\citen{Pattersons2021}], given by $\langle R \rangle = \frac{R_{\text{POL}} + R_{\text{EQ}}}{2}$. In this context, investigating the canonical mass $M=1.4\:M_\odot$ is particularly relevant, as observational data suggest that NSs predominantly reside within a relatively narrow mass range around $1.4 M_\odot$ \cite{Kaper2006}. 

As shown in Appendix~\ref{app:macroprops}, variations in $\gamma$ affect the $\langle R \rangle$ at both the maximum mass and the canonical mass only at the order of $10^{-3}$, which represents the largest magnitude of influence found---and even then, only for a subset of the configurations. Most of the values remain unchanged. In fact, the maximum mass is entirely unaffected by variations in $\gamma$, except for the configuration with $\lambda = -0.6 \times 10^{-5}$ and $\Omega = 0$, where the maximum mass is 1.99~$M_\odot$ for $\gamma = 3$ and 1.98~$M_\odot$ for $\gamma = 5$. This indicates that the impact of $\gamma$ on the overall stellar structure is minimal within the explored parameter range. Similar results for the same parameter range have also been reported in Refs.~[\citen{Lopes2015,Lopes2020,Wu2017}]. The appearance of the magnetic field generally decreases the eccentricity, which means that generally, the deformation is weakened once the magnetic field appears.

In the context of Rastall parameter, the greater values of $\lambda$ result in more compact NSs as the radii decrease, in which this fact is obviously presented in Fig. \ref{fig:static}. The greater values of Rastall parameter also cause the eccentricity diminished, in which this result can be validated in Fig. \ref{fig:eccentricity}. {It is important to note that the Rastall parameter exerts only a negligible influence on the maximum mass of NSs. A comparable behavior, where modified gravity has little effect on the stellar mass—particularly on the maximum mass—has also been reported in Ref. [\citen{OdintsovOikonomou2023}] (see Figs. 4 and 5 therein), where the NS mass is expressed as the ADM mass within the modified gravity framework adopted in their analysis.}

In term of the rotation, as we expect, faster rotations result in the increment of the eccentricity, since the deformation becomes stronger as the rotation is getting faster.

{Fig. \ref{fig:static} presents the mass-radius relation of NSs in a static configuration. This relation purely comes from TOV equation. From the relation, we can see that the presence of a chaotic magnetic field can increase the maximum mass of the NSs and simultaneously reduce the radius of NSs, as indicated by the leftward shift of the curves with the appearance of magnetic field strength, suggesting that a chaotic magnetic field can contribute to greater compactness of the stars. Furthermore, greater values of Rastall parameter lead to the decrease of the radii of NSs which also directly lead to the increment of NSs' compactness. Compactness, defined as the mass-to-radius ratio, is directly influenced by this field. Moreover, the appearece of the chaotic magnetic field results in the mass of NSs reach the mass range of J0348+0432, i.e. $2.01\pm 0.04\:M_\odot$ \cite{Antoniadis2013}, shown by yellow rectangle}. All mass curves are also in the agreement with the radius range of the canonical mass 1.4 $M_\odot$, shown by light blue line. The radius range of the canonical mass is taken from Ref.~[\citen{Lopes2024}], which is based on the observational results reported in Refs.~[\citen{Riley2019,Miller2019}], i.e. $11.52$ km $<R_{1.4}<$ $14.26$ km.


{In terms of the shape of the mass-radius relation curves, particularly when varying $\gamma$ within the range of 3 to 5, our results remain in qualitative agreement with the aforementioned studies. The results presented in Refs.~[\citen{Lopes2015,Lopes2020}] exhibit behavior more similar to ours, where the differences are relatively minor and the curves appear to coincide. It is worth noting that Ref.~[\citen{Wu2017}] employed the same chaotic magnetic field ansatz. Although the same range of $\gamma$ produces curves that are close to each other, they do not coincide as closely as in our results and those of Refs.~[\citen{Lopes2015,Lopes2020}]. These variations may also arise from differences in the underlying EOS models, as it is well known that different EOS models can lead to different outcomes in neutron star structure calculations~\cite{Rather2021}.}


Fig.~\ref{fig:deform} shows the polar and equatorial radii, indicating the deformation of rotating NSs. The deformation is less pronounced for NSs with higher masses but becomes more evident for lower masses. Note that even when the magnetic field vanishes, both in the GR and RG configurations, the deformation still occurs. This rotational deformation is expected and is similar to the Earth’s equatorial bulge, where the polar radius is smaller than the equatorial radius \cite{Pattersons2021}. The key point is that the deformation is weakened by the presence of the magnetic field and greater values of~$\lambda$. Since the magnetic field is embedded within the energy density and pressure terms, which appear in the perturbative expansion of the HT formalism, it can influence the deformation of the rotating NSs. Likewise, $\lambda$ appears in the quadrupole sector of the HT formalism, thereby contributing to strengthening the deformation (in the case of negative $\lambda$) or weakening it (in the case of positive $\lambda$).

We present 3- and 2-dimensional illustrations of the NSs in Fig. \ref{fig:deformation}. We take examples from the case of {the configuration $\lambda=-1\times10^{-5}$, $\gamma=3$, $\Omega=2000$ s$^{-1}$ . At lower mass, the star's shape becomes oblate enough. At $M=0.35\:M_\odot$, $R_{POL}$ = 13.14 km, and $R_{EQ}$ = 16.34 km; at $M=0.78\:M_\odot$, $R_{POL}$ = 13.18 km, and $R_{EQ}$ = 14.36 km; at $M=1.40\:M_\odot$, $R_{POL}$ = 13.56 km, and $R_{EQ}$ = 14.18 km; and at $M=2.00\:M_\odot$, $R_{POL}$ = 12.43 km, and $R_{EQ}$ = 12.69 km.} It is noteworthy to compare our results with those presented in Ref.~[\citen{Pattersons2021}]. The deformation of rotating NSs in our study is significantly more pronounced than in their findings, which are based on rotating NSs with anisotropic pressure of matter, but without considering the contribution of a magnetic field to the deformation.

The oblateness of NSs can also be characterized by their eccentricity $e$. In the context of a two-dimensional body, when the value of $e$ approaches 0, the shape of the body tends to be circular. Conversely, as $e$ approaches 1, the shape becomes more elliptical. Fig.~\ref{fig:eccentricity} illustrates the relationship between mass and eccentricity. Rotation, chaotic magnetic fields, and the Rastall parameter contribute to an increased deformation of NSs, as higher values of $\Omega$ drive the eccentricity $e$ closer to 1, while larger values of the Rastall parameter and the magnetic fields generally lead to a decrease in the eccentricity, as previously discussed when addressing Appendix~\ref{app:macroprops}.




Fig. \ref{fig:momin} depicts the relationship between the mass of NSs in static configuration and the moment of inertia of NSs and the corresponding eccentricity of the NSs. {It is worth noting that the Eqs. (\ref{angularmomentum}) and (\ref{momenofinersia}) that help us to calculate the moment of inertia of NSs are only valid for spherical shape. So, to keep the validity of the spherical-body approximation, we set the angular velocity to $\Omega = 50$ s$^{-1}$. As shown in Fig. \ref{fig:momin}(b), the resulting eccentricity values are found to be very close to zero, confirming that the stars remain nearly spherical under this condition. This justifies the continued use of the standard moment of inertia formula. Similar treatments for very slowly rotating NSs have also been employed in Refs.~[\citen{Rahmansyah2020,Rizaldy2024}], thereby providing precedent and support for our approach}.



We can see that at lower masses, the curves of moment of inertia tend to coincide, suggesting that the impact of chaotic magnetic field strength on the moment of inertia is negligible in this regime. However, a noticeable separation appears within the mass range $M=1.5-1.99$ $M_\odot$, where the appearance of chaotic magnetic fields lead to an increment in the moment of inertia. On the other hand, we also can see Rastall parameter has little impact on the moment of inertia of the NSs.

Remarkably, our findings for the moment of inertia are consistent with the constraint range provided by Landry et al. \cite{Landry2020}. It is worth noting that this constraint evaluates whether rotating NSs with chaotic magnetic fields align with observational evidence. The constraint is given as $I=1.78^{+0.66}_{-0.66}\times10^{45}$ g cm$^2$ at $M=1.4$ $M_\odot$, the standard canonical mass of NSs. As detailed in Ref.~[\citen{Landry2020}], this constraint is derived from radio observations of massive pulsars.

\section{Conclusion}
{In this study, we extend the formulation of rotating NSs with chaotic magnetic field in RG theory, by considering the effect of Rastall parameter in the quadrupole sector of HT formalism. We calculated the mass and radius of NSs under the influence of a chaotic magnetic field within GR and RG framework, enabling us to obtain the mass-radius relationship for these stars. The magnetic field ansatz used in this work is the one that is coupled to the energy density, which is proposed in Ref.~[\citen{Lopes2015}]. This ansatz behaves as parameter-free when $\gamma \geq 2$. We use three configurations: (i) $B = 0$; (ii) $B_0 = 3 \times 10^{18}$~G and $B_{\text{surf}} = 10^{15}$~G with $\gamma = 3$; and (iii) $B_0 = 3 \times 10^{18}$~G and $B_{\text{surf}} = 10^{15}$~G with $\gamma = 5$. In the context of RG, we use 5 values of Rastall parameter $\lambda$, i.e. $\lambda=-1\times10^{-5}$, $\lambda=-0.6\times10^{-5}$, $\lambda=0$ (GR), $\lambda=0.6\times10^{-5}$, and $\lambda=1\times10^{-5}$. For the rotation, we employ three values of $\Omega$, i.e. $\Omega=0$ (static configuration), $\Omega=1000$ s$^{-1}$, and $\Omega=2000$ s$^{-1}$.

Our findings indicate that both magnetic fields and greater values of Rastall parameter can decrease the radius of NSs. NSs formed with chaotic magnetic fields exhibit a higher maximum mass compared to those without chaotic magnetic fields. Additionally, the chaotic magnetic fields and greater values of Rastall parameter can contribute to a reduction in radii. This fact shows that both the chaotic magnetic fields and larger values of Rastall parameter can simultaneously enhance the compactness. Moreover, both factors weaken the deformation of NSs. Thus, chaotic magnetic fields and greater values of Rastall parameter decrease the values of eccentricity of the stars. In term of angular velocity $\Omega$, higher value of $\Omega$ strengthens the stellar deformation, as we expect. For deformed rotating NSs, significant deformations occur at lower mass values.}

{On the other side, our results indicate that the presence of chaotic magnetic field enhances the NSs' moment of inertia, in which the impact is obvioulsy appear in the mass range $1.50-1.99 M_\odot$. Nevertheless, the Rastall parameter does not significantly impact the moment of inersia of the NSs. Additionally, the moment of inertia of rotating NSs in our study is consistent with the constraint range obtained from adio observations of massive pulsars}

\section*{Acknowledgments}
We sincerely thank Prof. Luiz L. Lopes for sharing the EOS data. We also acknowledge Tsuyoshi Miyatsu and Ryan Rizaldy for valuable discussions. We are grateful to Anna Campoy Ordaz for making her code publicly available, which allowed MLP to modify it for use in this work. MLP acknowledges the Indonesia Endowment Fund for Education (LPDP) for financial support. FPZ and GH would like to thank Kemendiktisaintek (The Ministry of Higher Education, Science, and Technology) of the Republic of Indonesia (through LPPM ITB, Indonesia) for partial financial support. HLP would like to thank the National Research and Innovation Agency (BRIN) for
its financial support through the Postdoctoral Program. HLP and MLP would like to thank the members of the Theoretical Physics Groups at Institut Teknologi Bandung for their hospitality. MFARS is supported by the Second Century Fund (C2F), Chulalongkorn University, Thailand. We thank the anonymous reviewers for their helpful comments and suggestions.

\section*{Declaration of generative AI and AI-assisted technologies in the writing process}

During the preparation of this work, M. Lawrence Pattersons used ChatGPT (OpenAI) in order to improve the language and readability of the manuscript. After using this tool, he reviewed and edited the content as needed and take full responsibility for the content of the publication.

\appendix
\section*{Appendix: Macroscopic properties of neutron stars under different configurations.}
\label{app:macroprops}
{The quantities are shown in the following table.}

\includepdf[pages=-,landscape=true,fitpaper=true]{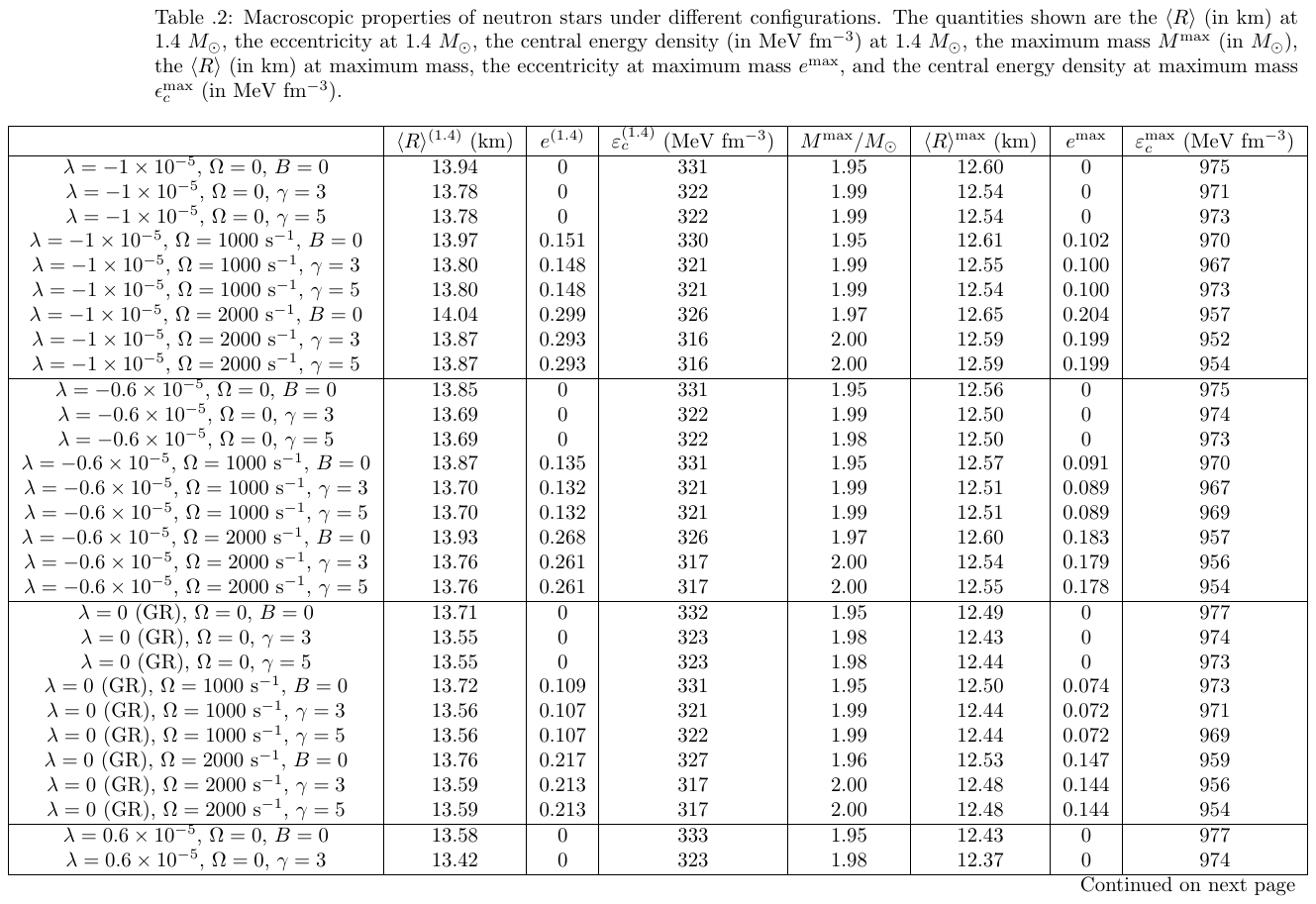}

\end{document}